\documentclass[a4paper,12pt]{article}
\usepackage{a4wide}
\usepackage{amsmath}
\usepackage{amsfonts}
\usepackage{bbm}
\usepackage{graphics}
\usepackage{graphicx}
\usepackage{subfigure}
\usepackage{pstricks}
\newcommand{\e}{\mathcal{E}}
\newcommand{\be}{\begin{eqnarray}}
\newcommand{\beq}{\begin{eqnarray}}
\newcommand{\ee}{\end{eqnarray}}
\newcommand{\tx}{\textrm}
\newcommand{\ov}{\overline}
\newcommand{\p}{\partial}

\newcommand{\cx}{\mathcal}
\newcommand{\Ga}{\Gamma}
\newcommand{\ab}{\ov{a}}
\newcommand{\ib}{\ov{i}}
\newcommand{\jb}{\ov{j}}

\newcommand{\bb}[1][\overline]{1}

\newcommand{\Fg}{\mathcal{F}^{(g)}}

\newcommand{\ue}{\text{e}}
\newcommand{\h}{\frac{1}{2}}

\begin{document}

\setlength{\parindent}{0cm}
\setlength{\baselineskip}{1.5em}

\title{\bf Global Properties of Topological String Amplitudes and Orbifold Invariants}
\author{Murad Alim, Jean Dominique L\"ange and Peter Mayr\footnote{\tt{murad.alim,dominique.laenge,mayr@physik.uni-muenchen.de}}\\[20pt]
\it Arnold Sommerfeld Center for Theoretical Physics\\
\it Ludwig-Maximilians-University \\
\it Department of Physics \\
\it Theresienstr. 37, D-80333 M\"unchen, Germany}
\date{}
\maketitle

\vspace{-250pt}
\hfill{LMU-ASC 49/08}
\vspace{+250pt}

\begin{center}
{\bf Abstract}
\end{center}
We derive topological string amplitudes on local Calabi-Yau manifolds in terms of polynomials in finitely many generators of special functions. These objects are defined globally in the moduli space and lead to a description of mirror symmetry at any point in the moduli space. Holomorphic ambiguities of the anomaly equations are fixed by global information obtained from boundary conditions at few special divisors in the moduli space. As an illustration we compute higher genus orbifold Gromov-Witten invariants for $\mathbbm{C}^3/\mathbbm{Z}_3$ and $\mathbbm{C}^3/\mathbbm{Z}_4$.

\clearpage

\section{Introduction}
In this paper we apply the method proposed in \cite{Yau,AL} to compute higher genus topological string amplitudes for local Calabi-Yau manifolds in terms of polynomials of a finite number of generating functions. The polynomial expression is globally defined and allows for an expansion of the topological string amplitudes at different points in moduli space. In particular we compute orbifold Gromov-Witten (GW) invariants which were recently studied on the physics and mathematics sides \cite{Aganagic,Bouchard,Brini,Coates}\footnote{We refer to \cite{Bouchard} for a complete list of references.} and make predictions for higher genus orbifold invariants for $\mathbbm{C}^3/\mathbbm{Z}_4$. 
\\
In \cite{AL}, it was shown that the non-holomorphic parts of the topological string amplitudes for any Calabi-Yau (CY) target space can be written as polynomials of a certain degree in a finite number of generators. This fact was proven by Yamaguchi and Yau \cite{Yau} for the quintic as well as other one parameter CYs and relies on the holomorphic anomaly equations of Bershadsky, Cecotti, Ooguri and Vafa (BCOV) \cite{BCOV1,BCOV2}. These equations relate the anti-holomorphic derivative of the genus $g$ topological string amplitudes $\mathcal{F}^{(g)}$ with holomorphic derivatives of amplitudes of lower genus. The equations provide thus recursive information for the full topological string partition function $Z=\exp(\sum \lambda^{2g-2}\mathcal{F}^{(g)})$. BCOV proposed further how to use special geometry to recursively determine $\mathcal{F}^{(g)}$ by partial integration once lower genus amplitudes are known. The polynomial method of \cite{Yau,AL} provides an improvement in doing this computation. The polynomial part is fixed recursively at every genus once the lower genus polynomials are known. 

The deficiency of the recursive information contained in the anomaly equation lies in the additional purely holomorphic data which have to be determined by boundary conditions \cite{BCOV2}. Knowledge of singularities and regularity of the amplitudes at various expansion points in moduli space has been pushed forward in \cite{HKQ} and was used together with Yamaguchi and Yau's polynomials to compute partition functions for the quintic to very high genus.
\\
The aim for our study is on one hand to understand details of the polynomial structure and the boundary conditions and on the other hand to demonstrate the power of this framework for computations of mathematical invariants at various expansion points in moduli space.\footnote{See \cite{Grimm} for other approaches adressing this problem.}
The outline of the paper is as follows. In the next section we review the results of \cite{AL}, discuss in detail the construction of the polynomial generators and analyze the freedom one has in choosing these. We identify the relevant boundary conditions at various expansion points and discuss how to use the fact that the polynomial expressions hold everywhere in moduli space to extract information. Then we apply this formalism to some well studied local Calabi--Yau threefolds, namely local $\mathbbm{P}_2$ which contains a $\mathbbm{C}^3/\mathbbm{Z}_3$ orbifold and local $\mathbbm{P}_1\times \mathbbm{P}_1$ as well as local $\mathbbm{F}_2$. The latter contains a $\mathbbm{C}^3/\mathbbm{Z}_4$ orbifold.
\\
At the time of finishing this work a paper appeared \cite{HKR} where the authors also use the polynomial construction to study local models.

\section{Polynomial Method}
\subsection{Review of the Polynomial Structure of Topological Strings}
In this section we review some of the results of \cite{AL} which are a generalization of \cite{Yau}.\\
The deformation space $\mathcal M$ of topological string theory, parameterized by coordinates
$z^i$, $i=1,...,\textrm{dim}(\mathcal{M})$, carries the structure of a special K\"ahler manifold.\footnote{See section two of \cite{BCOV2} for background material.} The ingredients of this structure are the Hodge line bundle $\mathcal{L}$ over $\mathcal M$ and the cubic couplings which are a holomorphic section of  $\mathcal{L}^{2} \otimes \textrm{Sym}^3 T^* \mathcal{M}$. The metric on $\mathcal{L}$ is denoted by $e^{-K}$ with respect to some local trivialization and provides a K\"ahler potential for the special K\"ahler metric on $\mathcal{M}$, $G_{i \jb}=\p_i \bar{\p}_{\jb}K $ . Special geometry further gives the following expression for the curvature  of $\mathcal{M}$ 
\begin{equation}
R_{i\ib\phantom{l}j}^{\phantom{i\ib}l}=[\bar{\p}_{\ib},D_i]^l_{\phantom{l}j}=\bar{\partial}_{\bar{i}} \Gamma^l_{ij}= \delta_i^l
G_{j\bar{i}} + \delta_j^l G_{i\bar{i}} - C_{ijk} C^{kl}_{\bar{i}}.
\label{curvature}
\end{equation}
The topological string amplitude or partition function $\Fg$ at genus $g$ is a section of the line bundle
$\mathcal{L}^{2-2g}$ over $\mathcal M$. The correlation function at genus $g$ with $n$
insertions $\mathcal{F}^{(g)}_{i_1\cdots i_n}$ is only non-vanishing for
$(2g-2+n)>0$. They are related by taking covariant derivatives as this
represents insertions of chiral operators in the bulk, e.g.~$D_i \mathcal{F}^{(g)}_{i_1\cdots i_n}=\mathcal{F}^{(g)}_{ii_1\cdots i_n}$.

Here $ D_i=\partial_i+\dots =\frac{\partial}{\partial z_i}+ \dots $
denotes the covariant derivative on the bundle $\cx L^m\otimes 
\tx{Sym}^n T^*\mathcal{M}$ where $m$ and $n$ follow from the context. $T^*\mathcal{M}$ is
the cotangent bundle of $\cx M$ with the standard connection coefficients
$\Ga^i_{jk}=G^{i\ib}\p_jG_{k\ib}$. The connection on the bundle $\cx L$ is
given by the first derivatives of the K\"ahler potential $K_i=\p_iK$.

Furthermore, in \cite{BCOV2} it is shown that the genus $g$ partition
function is recursively related to lower genus partition
functions. This is expressed  for $(2g-2+n)>0$ by the holomorphic
anomaly equations
\begin{equation}
\bar{\partial}_{\bar{i}} \Fg = \h \bar{C}_{\bar{i}}^{jk} \left(
\sum_{g_1+g_2=g}
D_j\mathcal{F}^{(g_1)} D_k\mathcal{F}^{(g_2)} +
D_jD_k\mathcal{F}^{(g-1)} \right) \label{feyn3},
\end{equation}
where
\begin{equation}
\bar{C}_{\bar{k}}^{ij}= \bar{C}_{\bar{i} \bar{j}\bar{k}} G^{i
\bar{i}}G^{j \bar{j}}\, \ue^{2K}, \qquad \bar{C}_{\bar{i}\bar{j}\bar{k}}=
\overline{C_{ijk}}\,. \label{feyn4}
\end{equation}
These equations, supplemented by \cite{BCOV1}
\begin{equation}
\bar{\partial}_{\bar{i}} \mathcal{F}^{(1)}_j = \frac{1}{2} C_{jkl}
C^{kl}_{\bar{i}}+ (1-\frac{\chi}{24})
G_{j \bar{i}}\,, \label{feyn5}
\end{equation}
and special geometry, determine all correlation functions up
to holomorphic ambiguities. In Eq.~(\ref{feyn5}), $\chi$ is the Euler
character of the manifold. A solution of the
recursion equations is given in terms of Feynman rules \cite{BCOV2}.

The propagators $S$, $S^i$, $S^{ij}$ for these Feynman rules are related
to the three point couplings $C_{ijk}$ as
\begin{equation}
\partial_{\bar{i}} S^{ij}= \bar{C}_{\bar{i}}^{ij}, \qquad
\partial_{\bar{i}} S^j = G_{i\bar{i}} S^{ij}, \qquad
\partial_{\bar{i}} S = G_{i \bar{i}} S^i.
\label{feyn7}
\end{equation}
By definition, the propagators $S$, $S^i$ and $S^{ij}$ are sections of
the bundles
$\mathcal{L}^{-2}\otimes \text{Sym}^m T$ with $m=0,1,2$.
The vertices of the Feynman rules are
given by the correlation functions $\mathcal{F}^{(g)}_{i_1\cdots
i_n}$.

The anomaly equation Eq.~(\ref{feyn3}), as well as the definitions
in Eq.~(\ref{feyn7}), leave the freedom of adding holomorphic
functions under the $\ov{\partial}$ derivatives as integration
constants. This freedom is referred to as holomorphic ambiguities.

The initial data for starting the recursion are the first non-vanishing correlation functions. At genus zero these are
the holomorphic three-point couplings $\mathcal{F}^{(0)}_{ijk} = C_{ijk}$.
The holomorphic anomaly equation Eq.~(\ref{feyn5}) can be easily integrated
with Eq.~(\ref{feyn7}) to
\begin{equation}
\mathcal{F}^{(1)}_i = \h C_{ijk} S^{jk} +(1-\frac{\chi}{24}) K_i
+ f_i^{(1)}, \label{sol2}
\end{equation}
with ambiguity $f_i^{(1)}$. As can be seen from
this expression, the non-holomorphicity of the correlation functions
only comes from the propagators together with $K_i$, in the following
called generators. Indeed, in
\cite{AL} it was proven by induction that this also holds for all partition
functions $\Fg$, $g>1$. One crucial step in this proof is that
the integrated special geometry relation
\begin{equation}
\Gamma^l_{ij} = \delta_i^l K_j + \delta^l_j K_i - C_{ijk} S^{kl} + s^l_{ij}\,,
\label{finite6}
\end{equation}
where $s^l_{ij}$ denote holomorphic functions that are not fixed by the
special geometry relation, allows to express the Christoffel connection in
terms of the generators. A second important step is to show that the covariant
derivatives of all generators are again expressed in terms of the generators \cite{AL}:
\begin{eqnarray}
D_i S^{jk} &=& \delta_i^j  S^k + \delta_i^k S^j - C_{imn}S^{mj} S^{nk} +
h_i^{jk} \label{finite9}, \\
D_i S^j &=& 2 \delta_i^j S - C_{imn} S^m S^{nj} + h_i^{jk} K_k  + h_i^j,
\label{finite10} \\
D_i S &=& -\frac{1}{2} C_{imn} S^m S^n + \frac{1}{2} h^{mn}_i K_m K_n +
h_i^j K_j + h_i, \label{finite11} \\
D_i K_j &=& -K_i K_j - C_{ijk} S^k + C_{ijk} S^{kl} K_l + h_{ij},
\label{finite12}
\end{eqnarray}
where $h_i^{jk}, h^j_i$, $h_i$ and $h_{ij}$ denote
holomorphic functions.
Assuming linear independence of $\bar{C}^{jk}_{\bar{i}}$ and $G_{i\bar{i}}$, the holomorphic anomaly equations
split into two equations
\begin{eqnarray}
\frac{\partial \Fg}{\partial S^{ij}} &=&  \h
\sum_{g_1+g_2=g}
D_i\mathcal{F}^{(g_1)} D_j\mathcal{F}^{(g_2)} + \h
D_iD_j\mathcal{F}^{(g-1)}, \label{rec2}
\\
0 &=& \frac{\partial \Fg}{\partial K_i} + S^i \frac{\partial \Fg}{\partial
S} + S^{ij} \frac{\partial \Fg}{\partial S^j}. \label{rec4}
\end{eqnarray}
The last Eq.~(\ref{rec4}) can be rephrased as the condition that
$\Fg$ does not depend explicitly on $K_i$ by making a suitable change of
generators
\begin{eqnarray}
\tilde{S}^{ij} &=& S^{ij}, \label{rec5} \\
\tilde{S}^i &=& S^i - S^{ij} K_j, \label{rec6} \\
\tilde{S} &=& S- S^i K_i + \frac{1}{2} S^{ij} K_i K_j, \label{rec7}\\
\tilde{K_i} &=& K_i, \label{rec10}
\end{eqnarray}
i.e.~$\partial \Fg / \partial \tilde{K}_i =0$ for $\Fg$ as a function
of the tilded generators. Finally, a $U(1)$ grading 
$1,2,3$ is assigned to the generators $S^{ij}$,
 $S^i$, $S$, respectively. The correlation functions
$\Fg_{i_1\cdots i_n}$ are polynomials of degree $3g-3+n$ in the
generators \cite{AL}. This finishes our review and we proceed with
some practical comments.
\subsection{Constructing the Propagators}
The application of the polynomial method requires the construction of the propagators and determination of the holomorphic functions that appear in Eqs.~(\ref{finite6}-\ref{finite12}). It should be noted that the discussion in the following holds in general and can be applied both for compact and local models. First one starts by defining the propagators $S^{ij}$, using the special geometry relation Eq.~(\ref{finite6}). We pick therefore a coordinate $z_{*}$ for which $C_{*jk}$ is invertible as a $n \times n$ matrix, this yields:
\begin{equation}
  S^{kl}  = (C_*^{-1})^{kj}( \delta_*^l K_j + \delta^l_j K_* -\Gamma^l_{*j}+ s^l_{*j})\;.
\label{propdef}
\end{equation}
We begin by analyzing the freedom in the definition of the propagators which is related to a choice of holomorphic $s^{l}_{ij}$ which have to satisfy  constraints coming from the symmetry of $S^{kl}$ and from the special geometry relations Eq.~(\ref{finite6}) with $i\ne *$. A counting of components of $s^l_{ij}$ minus constraints gives
\begin{equation}
\underbrace{\frac{n^2(n+1)}{2}}_{\textrm{components of } s_{ij}^l}- \underbrace{\frac{n(n-1)}{2}}_{\textrm{symmetry of } S^{ij}}- \underbrace{\frac{n^2(n-1)}{2}}_{\textrm{remaining special geometry}}= \frac{n(n+1)}{2}\;.
\end{equation}
This is equal to the number of components of a symmetric holomorphic $\e^{ij}$. This $\e^{ij}$ can be added to $S^{ij}$  while still satisfying the defining requirement $\partial_{\bar{i}} S^{ij}= \bar{C}_{\bar{i}}^{ij}$. Two choices $\tilde{s}_{ij}^l$ and $s_{ij}^l$ are related by:
\begin{equation}
\tilde{s}_{ij}^l = C_{ijk} \e^{kl} + s_{ij}^l\;.
\end{equation}
In the next step we tackle Eq.~(\ref{finite9}) where it is obvious that $h_{i}^{jk}$ for ($i\ne j$ and $i\ne k$) can already be computed with no freedom left, also the differences $h_i^{ii}-2 h_j^{ji}$ for $i \ne j$ can be computed. This leaves us with $n$ free holomorphic $h_i^{ii}$. These are related to a freedom in $S^i$ which we can define from Eq.~(\ref{finite9})
\begin{equation}
S^i=  \frac{1}{2} \left( D_i S^{ii}  + C_{imn}S^{mi} S^{ni} -
h_i^{ii} \right)\,. \label{propdef2}
 \end{equation}
Moving on to Eq.~(\ref{finite10}) we can now compute $h^i_j$ with $i \ne j$ and we obtain $n-1$ relations $h_i^i=h^j_j$ for $i \ne j$. This leaves the freedom to choose just one holomorphic $h^i_i$. Again this is related to a freedom in $S$ which can be defined from Eq.~(\ref{finite10})
\begin{equation}
S= \frac{1}{2} \left(D_i S^i+ C_{imn} S^m S^{ni} - h_i^{ik} K_k  - h_i^i \right)\,.
\label{propdef3}
\end{equation}
The remaining $h_{ij}$ in Eq.~(\ref{finite12}) can now be computed from the choices already made. This whole analysis of freedom in defining the propagators shows that given a set of propagators one can always add holomorphic pieces to each one. Of course holomorphic shifts in $S^{ij}$ affect $S^i$ and $S$. 
\subsubsection*{\it Holomorphic Freedom and Simplification}
Choosing a set of propagators amounts thus to choosing their holomorphic parts, their non-holomorphic part is fixed by the defining Eqs.~(\ref{feyn7}). The freedom one has in the construction can be summarized as follows
\begin{eqnarray}
S^{ij} &\rightarrow& S^{ij} + \mathcal{E}^{ij} \;,\\
S^i &\rightarrow& S^i + \mathcal{E}^{ij} K_j + \mathcal{E}^i\;, \\
S &\rightarrow& S + \h \mathcal{E}^{ij} K_i K_j + \mathcal{E}^i K_i +
\mathcal{E}\;,
\end{eqnarray}
all the holomorphic quantities in the polynomial setup change accordingly.

So $\e^{ij}, \e^i$ and $\e$ contain all the freedom in this polynomial setup. One special choice of propagators consistent with the Eqs.~(\ref{finite9}-\ref{finite12}) is such that their holomorphic part vanishes. In that case, given the holomorphic limit of the connection parts as analyzed by BCOV \cite{BCOV2} 
\begin{equation} \label{hollimit}
 K_{\textrm{hol}}= -\log X^0, \quad (\Gamma^k_{ij})_{\textrm{hol}}= \frac{\p z^k}{\p t^a} \frac{\p^2 t^a}{\p z^i \p z^j}\;,
\end{equation}
 where $t^a$ denote special\footnote{Or, more generally, canonical \cite{BCOV2}.} coordinates and $X^0$ the period used to define these. All the holomorphic quantities in the equations become trivial apart from $s_{ij}^l$ of Eq.~(\ref{finite6}) and $h_{ij}$ of Eq.~(\ref{finite12}) which are expressed by
\begin{eqnarray} \label{sandh}
 s_{ij}^l &=& \frac{\p z^l}{\p t^a} \frac{\p^2 t^a}{\p z^i \p z^j} + \delta_i^l \p_j \log{X^0} + \delta^l_j \p_i \log{X^0} , \nonumber\\
h_{ij} &=& -\p_i \p_j \log X^0 + \frac{\p z^m}{\p t^a} \frac{\p^2 t^a}{\p z^i \p z^j} \, \p_m \log{X^0} + \p_i\log X^0  \,\p_j \log X^0 .
\end{eqnarray}
This choice is of little use however as the polynomial part of the topological string amplitudes would be zero in the holomorphic limit and all the interesting information which allows to compute invariants would be encoded in the ambiguity. The goal is thus to use the insights of the polynomial structure coming from the non-holomorphic side of topological string theory to organize the amplitudes in a tractable form. In particular this means that we will try to absorb all the nontrivial series appearing in the holomorphic limit inside the holomorphic part of the propagators so that all remaining purely holomorphic quantities are simple closed expressions. Having done that the holomorphic ambiguity at every genus becomes a simple closed expression. 
\\
The periods giving $t^a$ and $X^0$ are computed patch-wise in the moduli space of complex structures by solving the Picard-Fuchs system of differential equations. In general the solutions are given in terms of series which have a finite radius of convergence and hence $s_{ij}^l$ and $h_{ij}$ will not have a closed form as can be seen from Eq.~(\ref{sandh}). Our guideline for choosing the holomorphic part of the propagators is to keep these expressions simple. We note that there are cases where it is possible to both have propagators that vanish in the holomorphic limit \emph{and} simple expressions for $s_{ij}^l$ and $h_{ij}$, namely in the cases where one has a:
\begin{itemize}
\item{\it Constant Period}\\
In local models the constant period is a solution of the Picard-Fuchs system, accordingly the holomorphic limit of derivatives of the K\"ahler potential vanish. If we take  Eq.~(\ref{finite12}) as a definition for the propagator $S^i$
\begin{equation}
S^k= (C_*^{-1})^{jk} (-D_* K_j -K_* K_j  + C_{*jk} S^{kl} K_l + h_{*j}),
\end{equation}
we see from Eq.~(\ref{hollimit}) that it is natural to choose its holomorphic limit to be zero which leads to vanishing $h_{ij}$. In this case the holomorphic limit of the remaining propagator $S$ vanishes too. This fact is referred to in the literature as triviality of these propagators for local models. We emphasize however that this actually only means that it is natural to choose the \emph{holomorphic limit} of these propagators to be zero. The propagators are not zero as they still are local potentials for the anti-holomorphic Yukawa couplings according to the definition Eq.~(\ref{feyn7}). The full topological string amplitudes expressed in terms of polynomials would still contain these quantities. Doing the computations in order to fix the ambiguity and to extract A-model invariants however does not require keeping track of these propagators.

\item {\it Special Mirror Map} \\
In the two parameter local models that we study we encounter the situation where one of the mirror maps just depends on one of the two parameters. From Eq.~(\ref{hollimit}) we see that in that case the holomorphic limit of the Christoffel connection with mixed lower indices vanishes. The latter appears in the definition of the propagators (\ref{propdef}) if we pick as $*$ the coordinate on which the mirror map does not depend. Accordingly a simple $s^k_{ij}$ leads to a trivial holomorphic limit for the corresponding propagator. In these cases we will later show that the corresponding propagators vanish identically and the models reduce effectively to one parameter models as noted before in \cite{Hosono} and in \cite{Aganagic}.
\end{itemize}

\subsection{Holomorphic ambiguity and boundary conditions}
To reconstruct the full topological string amplitudes we have to determine the purely holomorphic part of the polynomial. This holomorphic ambiguity can be fixed by imposing various boundary conditions at special points in the moduli space. 

The leading behavior at large complex structure\footnote{We will use the term ``large complex structure'' to denote the expansion point in the moduli space of the B-model which is mirror to large volume on the A-side.} was
computed in \cite{BCOV1,BCOV2,MM,GV,FP}. In particular the 
contribution from constant maps is 
\begin{equation} \label{constmaps}
 \mathcal{F}^{(g)}|_{q_a=0}= (-1)^g \frac{\chi}{2} \frac{|B_{2g} B_{2g-2}|}{2g\,(2g-2)\,(2g-2)!} \; , \quad g>1,
\end{equation}
where $q_a$ denote the exponentiated mirror maps at large radius.  

The leading singular behavior of the partition functions at a conifold locus has been determined in \cite{BCOV1,BCOV2,Ghoshal,Anton,GV}
\begin{equation} \label{Gap}
 \mathcal{F}^{(g)}(t_c)=  \frac{B_{2g}}{2g (2g-2) t_c^{2g-2}} +\cx O(t^0_c),
\qquad g>1
\end{equation}
Here $t_c\sim \Delta$ is the canonical coordinate 
at the discriminant locus $\Delta=0$ of a simple conifold. In particular the 
leading singularity in \eqref{Gap} as
well as the absence of subleading singular terms follows from the 
Schwinger loop computation of \cite{GV}, which computes the effect 
of the extra massless hypermultiplet 
in the space-time theory \cite{Vafacf}. The singular structure and the ``gap''  
of subleading singular terms have been also observed in the dual matrix model
\cite{Aganagic:2002wv} and were first used in \cite{HK,HKQ} 
to fix the holomorphic ambiguity to very high genus.

Note that the space-time derivation of \cite{GV} is
not restricted to the conifold case and applies also to Calabi--Yau 
singularities which give rise to a different spectrum of
extra massless vector and hypermultiplets in space-time. So more generally one expects 
the singular structure 
\begin{equation} \label{Gapp}
 \mathcal{F}^{(g)}(t_c)=  b \frac{B_{2g}}{2g (2g-2) t_c^{2g-2}} +\cx O(t^0_c),
\qquad g>1
\end{equation}
with $t_c\sim \Delta^\gamma$, $\gamma>0$. The coefficient of
the Schwinger loop integral is a weighted trace over the spin of the particles 
\cite{Vafacf, Anton} leading to the prediction $b=n_H-n_V$ for the coefficient of the leading singular term.
In section 3.3 we will 
consider an example with a singularity that gives rise
to a $SU(2)$ gauge theory in space-time and find agreement with the singular behavior and the generalized
gap structure predicted by the Schwinger loop integral.

\def\tz{\tilde{z}}
The singular behavior is taken into account by the local ansatz
\begin{equation}\label{ansatzha}
\mathrm{hol. ambiguity}\sim \frac{p(\tz_i)}{\Delta^{(2g-2)}},
\end{equation}
for the holomorphic ambiguity near $\Delta=0$, 
where $p(\tz_i)$ is a priori a series in the local coordinates $\tz_i$ near the singularity. Patching
together the local informations at all the singularities with the boundary 
divisors $z_i\to\infty$ for one or more $i$
 it follows however that the nominator 
$p(z_i)$ is generically a polynomial of low degree in the $z_i$. Here $z_i$ denote the natural coordinates centered at 
large complex structure, $z_i=0\ \forall i$.
Generically the $\mathcal{F}^{(g)}, g>1 $ are regular at 
a boundary divisor $z_i\to \infty $ 
from which it follows that the degree of $p(z_i)$ in $z_i$ is smaller or equal
to the maximum power of $z_i$ appearing in the discriminants 
in the denominator.%
\footnote{It can also happen, that a boundary divisor $z_i=\infty$ gives
rise to a singularity of the type \eqref{Gap}. For compact manifolds one has to take into account the effect of gauge transformations between different patches.} 

The finite number of coefficients in $p(z_i)$ is constrained by \eqref{Gapp}. In the computations 
for the local Calabi--Yau models considered below it turns out that the 
boundary conditions described above are sufficient to fix the holomorphic ambiguities.

\section{Application to Local Mirror Symmetry}
 Mirror symmetry in topological string theory refers to the equivalence of the A-model on a family of target spaces $X_t$ which are related by deformations of K\"ahler structure on one side and the B-model on the family of target spaces $Y_z$ which are related by deformations of complex structure on the other side. The mirror map $t(z)$, which represents the matching between the deformation spaces (moduli spaces) was first found at the large K\"ahler/Complex Structure expansion point \cite{Candelas}. This matching between the theories on both sides is believed to exist everywhere in the moduli space
which has generically different phases \cite{phases,Aspinwall2,Aspinwall3}. The precise matching between the A-model and the B-model has to be found for each point in the moduli \cite{BCOV2}. We will do this for special points in the case of some non-compact, i.e.~local models. 
Local mirror symmetry has been developed in \cite{KlemmMV,Engineering,KMV,Chiang,KlemmZaslow}. For reviews of many subsequent developments and a list of references see \cite{MarinoCS,MarinoMM}. 

The models we consider are described torically on the A-side. The A-side is most compactly described by giving the set of charge vectors $\vec{l}^{(a)}$ with $a=1,\dots,\textrm{dim} \,h^{(1,1)}(X)$. The number of components of each vector corresponds to the number of homogeneous coordinates on the toric variety. The B-model moduli space is described by the secondary fan which is obtained from the columns of the matrix of charge vectors. This description is very useful for obtaining the right coordinates describing each phase. For the polynomial construction we analyze the information that can be obtained from each phase. 
\subsection{Local $\mathbbm{P}^2$}
Local $\mathbbm{P}^2$ denotes the anti-canonical bundle over $\mathbbm{P}^2$, $\mathcal{O}(-3)\rightarrow \mathbbm{P}^2$,  which can be obtained by taking one K\"ahler parameter of a two parameter compact Calabi-Yau to infinity. The compact CY is a torus fibration which is described by the charge vectors
\begin{eqnarray}
l^{(a)}=\left(
\begin{array}{c|cccccc}
0&0&0&-3&1&1&1\\
-6&3&2&1&0&0&0
\end{array}\right)\,.
\end{eqnarray}
We will denote by $t_1$ and $t_2$ the K\"ahler parameters of the base and the fiber respectively. The limit $t_2 \rightarrow i \infty$ corresponds to the decompactification. To take this limit we must find a linear combination of dual periods $\mathcal{F}_{t_i}$ which remains finite in this limit \cite{Chiang}. This combination is\footnote{We analyzed the leading terms coming from classical intersection numbers which can be computed from the toric data.}
\begin{equation}
 (\p_{t_1} -\frac{1}{3} \p_{t_2}) \mathcal{F}=-\frac{1}{6}t_1^2+ \dots\, .  
\end{equation}
After the change of coordinates
\begin{equation}
 t=t_1\,, \quad s=t_2+\frac{1}{3} t_1\,,
\end{equation}
this dual period is rephrased as $\p_t \mathcal{F}^{(0)}= -\frac{1}{6}t^2+ \dots$, which can be integrated to give the prepotential of the local model. The ``classical intersection'' numbers in the basis given by $t$ and $s$ are then 
\begin{equation}
 C^{(0)}_{ttt}=-\frac{1}{3}\,, \quad C^{(0)}_{sss}=9\,,  \quad \textrm{and} \quad C^{(0)}_{sst}=C^{(0)}_{stt}=0\,.
\end{equation}
With the expression for the K\"ahler potential in special coordinates\footnote{$\mathcal{F}_{a}:=\frac{\p\mathcal{F}}{\p t_a}$.}
\begin{equation}
 e^{-K}= i |X^0|^2 \left( 2(\mathcal{F}-\ov{\mathcal{F}})-(t_a-\ov{t}_{\ab})(\mathcal{F}_{a}+\ov{\mathcal{F}}_{{\ab}}) \right),
\end{equation}
we find for the inverse $tt^*$ metric 
\begin{eqnarray}
g^{a\ov{b}}:= e^{K} G^{a\ov{b}}=\left(\begin{array}{cc}
g^{t\ov{t}}&g^{t\ov{s}}\\
g^{s\ov{t}}&g^{s\ov{s}}
\end{array}\right) \stackrel{s \rightarrow i\infty }{\longrightarrow} \frac{1}{2\textrm{Im} \tau} 
\left(\begin{array}{cc}
1&0\\
0&0
\end{array}\right), 
\end{eqnarray}
where $\tau:=\mathcal{F}_{tt}$. The $tt^*$ metric appears in the definition of the propagators, here in special coordinates
\begin{equation} \label{propspec}
\p_{\bar{a}} S^{bc}=\bar{C}_{\bar{a}}^{bc}= \ov{C}_{\bar{a}\bar{b}\bar{c}} g^{b \bar{b}} g^{c \bar{c}},
\end{equation}

which shows that in that limit the propagators $S^{ts}$  and  $S^{ss}$  vanish.

\begin{figure}[tbp]
\centering
\begin{minipage}{0.8\textwidth}
\centering
\subfigure[ A-model.]{
\centering
\psset{unit=2.3cm}
\begin{pspicture}(-1,-1)(1,1)
\psline[linewidth=1pt](-1,-1)(0,0)
\psline[linewidth=1pt](-1,-1)(0,1)
\psline[linewidth=1pt](-1,-1)(1,0)
\psline[linewidth=1pt](0,0)(0,1)
\psline[linewidth=1pt](0,0)(1,0)
\psline[linewidth=1pt](1,0)(0,1)

\rput[tr](-1.07,-1.03){$a_3$}
\rput[bl](0.07,1.03){$a_2$}
\rput[bl](1.03,0.07){$a_1$}
\rput[bl](0.07,0.07){$a_0$}

\psdot[dotsize=5pt 1,dotstyle=o](-1,-1)%
\psdot[dotsize=5pt 1,dotstyle=o](1,0)%
\psdot[dotsize=5pt 1,dotstyle=o](0,1)%
\psset{fillcolor=black}
\psdot[dotsize=5pt 1,dotstyle=o](0,0)%
\end{pspicture}
}\hfill
\subfigure[ B-model.]{
\centering
\raisebox{2.2cm}[0cm]{
\psset{unit=1.4cm}
\begin{pspicture}(-3,0)(1,0)
\psline[arrows=->,arrowsize=5pt,arrowinset=0](0,0)(-3,0)
\psline[arrows=->,arrowsize=5pt,arrowinset=0]{->}(0,0)(1,0)

\rput[l](1.07,0){$1$}
\rput[r](-3.07,0){$-3$}
\rput[br](1.1,0.11){$a_1,$ $a_2,$ $a_3$}
\rput[br](-2.5,0.11){$a_0$}

\psset{fillcolor=black}
\psdot[dotsize=3pt 0,dotstyle=o](0,0)%
\end{pspicture}}}
\caption{Fan and secondary fan for local $\mathbbm{P}^2$.}
\label{fig_p2}
\end{minipage}
\end{figure}
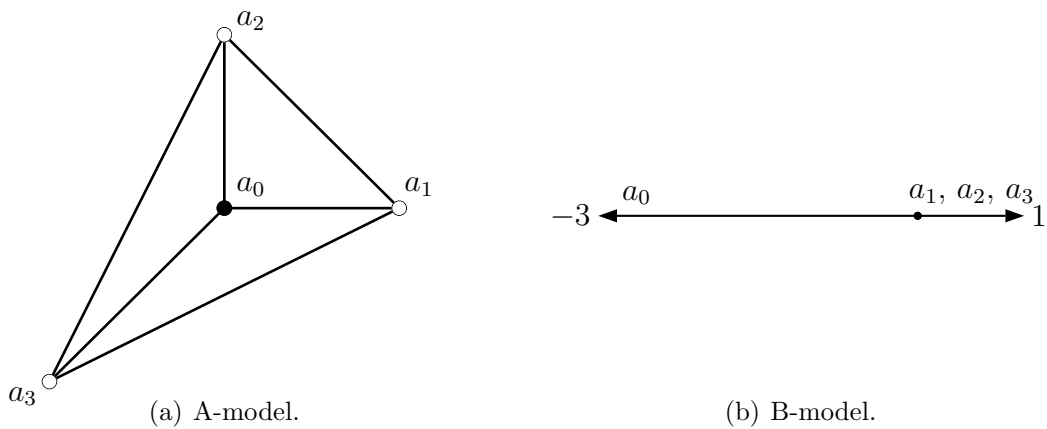
We can describe local $\mathbbm{P}^2$ torically by the charge vector $l=(-3,1,1,1)$. 

Local mirror symmetry associates to this non-compact toric variety a one dimensional local geometry \cite{Engineering,KMV} described by
\begin{equation}
P=a_0 u_1 u_2 u_3 + a_1 u_1^3+ a_2 u_2^3 +a_3 u_3^3 =0\;,
\end{equation}
where $u_1,u_2$ and $u_3$ are projective coordinates. The rescaling $u_i \rightarrow \lambda_i u_i$ induces a $(\mathbbm{C}^*)^3$ action on $a_i$
\begin{equation}
 (a_0,a_1,a_2,a_3)\rightarrow (\lambda_1\lambda_2\lambda_3 a_0,\lambda_1^3a_1,\lambda_2^3 a_2,\lambda_3^3 a_3)\,.
\end{equation}
Invariant combinations of $a_i$ under this action parameterize the moduli space of complex structures of the mirror geometry. Different phases on the B-side are encoded in the secondary fan which is one dimensional and shows that the moduli space of complex structures has two patches. These correspond on the A-model side to the $\mathbbm{C}^3/\mathbbm{Z}_3$ orbifold phase and to the blown up phase. The Picard Fuchs system in homogeneous coordinates is given by\footnote{$\theta_{a}:=a\frac{\p}{\p a}$.}
\begin{equation}
 \mathcal{L} = \theta_{a_1} \theta_{a_2} \theta_{a_3}-  \frac{a_1a_2a_3}{a_0^3}\theta_{a_0} (\theta_{a_0}-1)(\theta_{a_0}-2)\;.
\end{equation}
We refer to the literature for a complete discussion of solutions of the Picard-Fuchs system for local $\mathbbm{P}^2$, see for example \cite{Diaconescu}. We will rather focus on the ingredients that we use for applying the polynomial method. We fix the choices of the holomorphic quantities in the polynomial procedure at large complex structure.
\subsubsection*{\it Large Complex Structure}
We use the induced $(\mathbbm{C}^*)^3$ action on $a_i$ to set $a_2,a_3,a_0\rightarrow 1, a_1\rightarrow \frac{a_1 a_2 a_3}{a_0^3}$.
The good coordinate in this patch is hence given by $ z= \frac{a_1a_2a_3}{a_0^3}$.
Our choice of data for Eqs.~(\ref{finite6}-\ref{finite12}) which completely fixes the polynomial construction is given by
\begin{equation}
 s_{zz}^z=-\frac{4}{3z}\,, \quad h_z^{zz}= -\frac{z}{3}\,, \quad \textrm{and} \quad h^z_z,h_z,h_{zz}=0\,.
\end{equation}
We further need the starting amplitudes of the recursion
\begin{eqnarray}
C_{zzz} &=&-\frac{1}{3 z^3} P\,, \quad \textrm{where} \quad P:=\frac{1}{\Delta}\;, \quad \Delta=1+27 z\,, \\
z \mathcal{F}^1_z &=& \frac{1}{12} (P-1)-\frac{1}{6} P \frac{S^{zz}}{z^2}+\frac{1}{12}\;.
\end{eqnarray}
We need to supplement the polynomial part with the holomorphic ambiguity at every genus, which we determine by moving to other patches in the moduli space of the B-model.
\subsubsection*{\it Orbifold}
We use the $(\mathbbm{C}^*)^3$ action to set $a_1,a_2,a_3\rightarrow 1, a_0 \rightarrow \frac{a_0}{(a_1a_2a_3)^{1/3}}$.
The good invariant coordinate to parameterize this patch is given by $x=\frac{a_0}{(a_1 a_2 a_3)^{(1/3)}}$, we find solutions for the Picard-Fuchs system
\begin{eqnarray}
\omega_0&=&1\,, \nonumber\\
\omega_1 &= &x - \frac{{x}^4}{648} + \frac{4\,{x}^7}{229635} - 
  \frac{49\,{x}^{10}}{159432300} + \dots\,, \nonumber\\
\omega_2 &=& {x}^2 - \frac{2\,{x}^5}{405} + \frac{25\,{x}^8}{367416} - \frac{160\,x^{11}}{122762871} + \dots\,. 
\end{eqnarray}
Monodromy around large complex structure $z\rightarrow e^{2\pi i} z$ induces a transformation of the solutions as $(\omega_0,\omega_1,\omega_2)\rightarrow  (\alpha^3 \omega_0, \alpha \omega_1,\alpha^2 \omega_2)$ with $\alpha=e^{\frac{-2\pi i}{3}}$.This describes an orbifold in the moduli space of the B-model and corresponds to a $\mathbbm{C}^3/\mathbbm{Z}_3$ target space on the A-side. Now we need to write these periods in the form $(1,t_o,\p_{t_o} \mathcal{F})$ such that $\mathcal{F}$ is monodromy invariant. This is possible by identifying $t_o=\omega_1$ and $\p_{t_o} \mathcal{F}= \frac{1}{6} \omega_2$. We introduced the factor $1/6$ in order to reproduce the genus zero Orbifold Gromov-Witten invariants that appeared in \cite{Aganagic,Bouchard}. Higher genus Orbifold invariants are only sensitive to the normalization of $t_o$ which we find to be $1$. Our actual purpose for moving to this patch is to examine the behavior of the polynomial and to restrict the ansatz we have to make at large complex structure for the holomorphic ambiguity. 

To do that we first note that the full expression for $\mathcal{F}^{(g)}$ does not transform under a coordinate change in the complex structure moduli space\footnote{They transform as sections of $\mathcal{L}^{(2-2g)}$ but in local models this transformation is trivial due to the constant period.}. To clarify this we look at a typical expression appearing in the BCOV Feynman graph expansion at genus $g$ which would have the form
$$ \mathcal{F}^{(g)}_{i_1\dots i_n} S^{i_1i_2}\dots S^{i_{n-1} i_n },$$
where all indices are contracted. Hence, the whole expression does not transform. In the polynomial formalism all quantities in $\mathcal{F}^{(g)}_{i_1\dots i_n}$ coming from the connections would also be expressed in terms of polynomial building blocks. We therefore only have to express the building blocks that we found at large complex structure in the new coordinates without worrying about tensor transformations of the indices. The only polynomial building block which survives when taking the holomorphic limit has the following expansion in $t_o$
\begin{equation}
\frac{S^{zz}}{z^2} (t_o) = \frac{1}{2}+\frac{1}{540} t_o^3 + \mathcal{O}(t_o^6)\,.
\end{equation}
We further check that the polynomial part at every genus has a regular expansion in $t_o$. The regularity of the polynomial expression depends on the choice of $s_{zz}^z$. Being sure that we have made an appropriate choice we can make an ansatz for the holomorphic ambiguity at large complex structure which is also regular at the orbifold and has the right singular behavior at the conifold,
\begin{equation}
 f^{(g)}(z)= \Delta^{2-2g}  \sum_{i=0}^{2g-2} a_{i} z^i.
\end{equation}
One of the $2g-1$ coefficients in the ansatz is determined by the contribution of constant maps at large radius Eq.~(\ref{constmaps}), the other $2g-2$ are determined by the gap condition at the conifold. To implement the gap condition we need to go to a patch where the coordinate is the discriminant.
\subsubsection*{ \it Conifold}
The coordinate on complex structure moduli space in this patch is given by
$$y=1+27 z\,.$$
We solve the Picard-Fuchs system in terms of $y$ and obtain the mirror map
\begin{equation}
t_c(y)= \frac{1}{\sqrt{3}} \left( y + \frac{11\,{y}^2}{18} + \frac{109\,{y}^3}{243} + 
  \frac{9389\,{y}^4}{26244} + \dots \right) \,.
\end{equation}
The normalization is such that $\mathcal{F}^{(g)}(t_c)$ has the behavior described in Eq.~(\ref{Gap}). Once the normalization is fixed a simple counting of conditions vs.~unknowns shows that the recursion is completely determined up to arbitrary genus. We list some low genus Orbifold Gromov-Witten invariants in the appendix.

\subsection{Local $\mathbbm{F}_0$}
Local $\mathbbm{F}_0$ denotes the anti-canonical bundle over $\mathbbm{P}^1 \times \mathbbm{P}^1$, which is obtained by taking one K\"ahler parameter of a three parameter compact Calabi-Yau to infinity. The compact CY is a torus fibration which is described by the charge vectors
\begin{eqnarray}
l^{(a)}=\left(
\begin{array}{c|ccccccc}
0&0&0&-2&1&1&0&0\\
0&0&0&-2&0&0&1&1\\
-6&3&2&1&0&0&0&0
\end{array}\right)\,.
\end{eqnarray}
We denote by $t_1$ and $t_2$ the K\"ahler parameters of the base and by $t_3$ the K\"ahler parameter of the fiber. The limit $t_3 \rightarrow i \infty$ corresponds to the decompactification. Again, to take this limit we must find a linear combination of dual periods $\mathcal{F}_{t_i}$ which remains finite in this limit. This combination is
\begin{equation}
 (\p_{t_1}+\p_{t_2} -\frac{1}{2} \p_{t_3}) \mathcal{F}=-\frac{1}{2}t_1 t_2+ \dots\,.  
\end{equation}
After the change of coordinates
\begin{equation}
 t=t_1\,, \quad u=t_2-t_1\,, \quad s=t_3+ \frac{1}{2} t_1\,,
\end{equation}
this dual period becomes
\begin{equation}
\p_t \mathcal{F} = -\frac{1}{2}t(u+t)+ \dots\,.
\end{equation}
Integrating this we obtain for the prepotential
\begin{equation}
 \mathcal{F}=-\frac{1}{6} t^3 -\frac{1}{4} t^2 u+\frac{a}{6} u^3 +\dots\,,
\end{equation}
where $a$ denotes an arbitrary constant which drops out in constructing the propagators but nevertheless affects the classical term when we determine the Yukawa couplings\footnote{This explains why different classical data for local models lead to the same results.}. We set this constant to zero.
The nonzero ``classical intersections'' in the new coordinates are
\begin{equation} \label{intersectionf0}
 C^{(0)}_{ttt}=-1\,, \quad C^{(0)}_{sss}=8\,,  \quad C_{uss}^{(0)}=2\,, \quad \textrm{and} \quad C^{(0)}_{ttu}=-\frac{1}{2}\,.
\end{equation}
We can redo the analysis of taking $s$ to infinity. We find in this case the following inverse $tt^*$ metric 
\begin{eqnarray}
g^{a\ov{b}}=\left(\begin{array}{ccc}
g^{t\ov{t}}&g^{t\ov{u}}&g^{t\ov{s}}\\
g^{u\ov{t}}&g^{u\ov{u}}&g^{u\ov{s}}\\
g^{s\ov{t}}&g^{s\ov{u}}&g^{s\ov{s}}
\end{array}\right) \stackrel{s \rightarrow i\infty }{\longrightarrow} \frac{1}{2 \textrm{Im} \mathcal{F}_{tt}}\,.
\left(\begin{array}{ccc}
1&0&0\\
0&0&0\\
0&0&0
\end{array}\right) 
\end{eqnarray}
This shows that all propagators containing $s$ vanish in that limit. More interestingly the propagators $S^{ut}$ and $S^{uu}$ also vanish. It was already observed in \cite{Hosono} that local $\mathbbm{F}_0$ is a two parameter problem which effectively reduces to a one parameter problem. In \cite{Aganagic} it was argued that only the K\"ahler parameters that correspond to 2-cycle classes which are dual to 4-cycles which remain compact are true parameters of the theory. This limit shows this from a different point of view. Only one propagator $S^{tt}$ survives the decompactification. It should be noted that we could still choose some non-zero holomorphic limit for the vanishing propagators as this analysis only shows the vanishing of their anti-holomorphic derivative, this would however be redundant.

Now we can examine which holomorphic anomaly equations survive the decompactification. We find that $\p_{\bar{t}} \Fg$ and $\p_{\bar{u}} \Fg$ give the same equation. As was proven in \cite{AL} the non-holomorphic dependence of $\Fg$ comes from the polynomial building blocks. The chain rule shows that the anomaly equation $\p_{\bar{u}} \Fg$ reduces to $\p_{\bar{t}} \Fg$. Only one non-trivial anomaly equation survives.\\
\begin{figure}[t!]
\centering
\begin{minipage}{0.7\textwidth}
\centering
\subfigure[ A-model.]{
\centering
\psset{unit=2.3cm}
\begin{pspicture}(-1,-1.2)(1,1)
\psline[linewidth=1pt](0,0)(1,0)
\psline[linewidth=1pt](0,0)(0,1)
\psline[linewidth=1pt](0,0)(-1,0)
\psline[linewidth=1pt](0,0)(0,-1)
\psline[linewidth=1pt](1,0)(0,1)
\psline[linewidth=1pt](0,1)(-1,0)
\psline[linewidth=1pt](-1,0)(0,-1)
\psline[linewidth=1pt](0,-1)(1,0)

\rput[tl](0.07,-1.03){$a_4$}
\rput[bl](0.07,1.03){$a_3$}
\rput[br](-1.03,0.07){$a_2$}
\rput[bl](1.03,0.07){$a_1$}
\rput[bl](0.07,0.07){$a_0$}

\psdot[dotsize=5pt 1,dotstyle=o](-1,0)%
\psdot[dotsize=5pt 1,dotstyle=o](1,0)%
\psdot[dotsize=5pt 1,dotstyle=o](0,-1)
\psdot[dotsize=5pt 1,dotstyle=o](0,1)%
\psset{fillcolor=black}
\psdot[dotsize=5pt 1,dotstyle=o](0,0)%
\end{pspicture}
}\hfill
\subfigure[ B-model.]{
\centering
\raisebox{0.5cm}[0cm]{
\psset{unit=1.4cm}
\psset{griddots=0,gridlabels=0,subgriddiv=5}
\begin{pspicture}(-2,-2)(1,1)
\psgrid[griddots=5,subgriddiv=0,gridcolor=gray]

\psline[arrows=->,arrowsize=5pt,arrowinset=0](0,0)(1,0)
\psline[arrows=->,arrowsize=5pt,arrowinset=0](0,0)(0,1)
\psline[arrows=->,arrowsize=5pt,arrowinset=0](0,0)(-2,-2)

\rput[l](1.07,0){$(1,0)$}
\rput[b](0,1.07){$(0,1)$}
\rput[t](-2.07,-2.07){$(-2,-2)$}
\rput[br](-1.85,-1.75){$a_0$}
\rput[br](0.8,0.09){$a_1,a_2$}
\rput[tl](0.07,0.8){$a_3,a_4$}
\rput[tr](0.9,0.9){I}
\rput[tl](0.4,-1.4){II}
\rput[br](-1.36,0.4){III}
\end{pspicture}}}
\caption{Fan and secondary fan for local $\mathbbm{F}_0$.}
\label{fig_f0}
\end{minipage}
\end{figure}
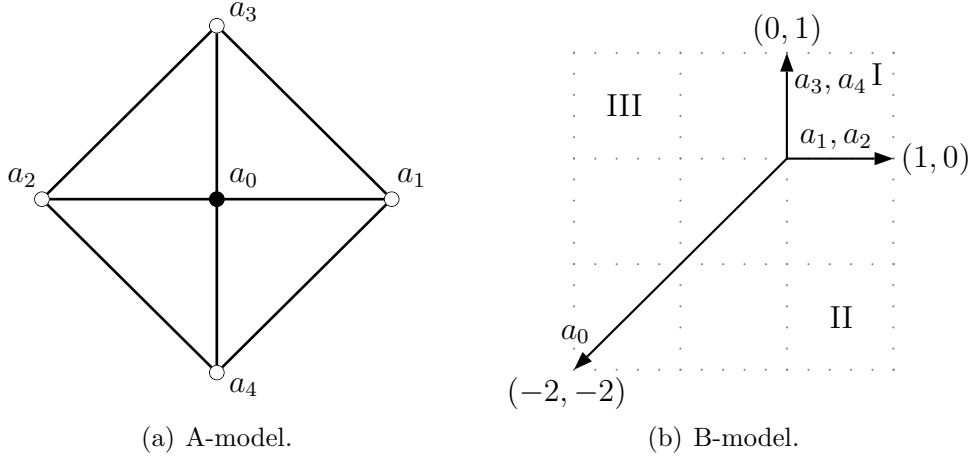
Local $\mathbbm{F}_0$ is described by the charge vectors
\begin{eqnarray}
l^{(a)}=\left(
\begin{array}{ccccc}
a_0&a_1&a_2&a_3&a_4\\
-2&1&1&0&0\\
-2&0&0&1&1
\end{array}\right)\,.
\end{eqnarray}
\\The mirror geometry is given by,
\begin{equation}
 P=a_0 u_1 u_2 u_3 u_4 + a_1 u_1^2 u_2^2 + a_2 u_3^2 u_4^2 +a_3 u_1^2 u_4^2+a_4 u_2^2 u_3^2=0\,.
\end{equation}
The rescaling $u_i \rightarrow \lambda_i u_i$, $\lambda_i \in
\mathbbm{C}^*$ induces a ${(\mathbbm{C}^*)}^3$-action
\begin{align}
(a_0,a_1,a_2,a_3,a_4) &\rightarrow (\lambda_1 \lambda_2 \lambda_3 \lambda_4
a_0,\lambda_1^2 \lambda_2^2 a_1,\lambda_3^2 \lambda_4^2 a_2,\lambda_1^2 \lambda_4^2
a_3, \lambda_2^2 \lambda_3^2 a_4)\,,
\end{align}
as only three rescalings are independent.
The Picard-Fuchs system is given by
\begin{eqnarray}
 \mathcal{L}_1 = \theta_{a_1}\theta_{a_2} - \frac{a_1 a_2}{a_0^2} \theta_{a_0}(\theta_{a_0}-1)\,, \nonumber\\
 \mathcal{L}_2 = \theta_{a_3}\theta_{a_4} - \frac{a_3 a_4}{a_0^2} \theta_{a_0}(\theta_{a_0}-1)\,.
\end{eqnarray}

\subsubsection*{\it Large Complex Structure, Region I}
We use the induced $\mathbbm({\mathbbm{C}^*})^3$ action to set $a_0,a_2,a_4\rightarrow 1, a_1 \rightarrow \frac{a_1 a_2}{a_0^2}$ and $a_3 \rightarrow \frac{a_3 a_4}{a_0^2}$. Good coordinates are therefore $z_1=\frac{a_1a_2}{a_0^2}$,
$z_2=\frac{a_3a_4}{a_0^2}$ with mirror maps\footnote{We absorb factors of $2\pi$ in the definition of $t_1,t_2$.}
\begin{align}
t_1 &= \log z_1 + 2 z_1 + 2 z_2 + 3 z_1^2 + 12 z_1 z_2 + 3 z_2^2
+ \cdots\,, \\
t_2 &= \log z_2 + 2 z_1 + 2 z_2 + 3 z_1^2 + 12 z_1 z_2 + 3 z_2^2
+ \cdots\,.
\end{align}
To make contact with our previous discussion we change $t_1$ and $t_2$ to $t=t_1$ and $u=t_2-t_1$ by an $SL(2,\mathbbm{Z})$ transformation, which changes the coordinates on the complex structure moduli space to
\begin{align}
y_1 = z_1\,, \quad y_2 = \frac{z_2}{z_1}\,,
\end{align}
indeed we now find a mirror map which only depends on $y_2$
\begin{align}
t(y_1,y_2)&= t_1 = \log y_1 + 2 y_1 + 3 y_1^2 + 2 y_1 y_2 + \cdots\,, \\
u(y_2)&= t_2-t_1 = \log y_2\,.
\end{align}
Calculating the holomorphic limit of the Christoffel connections we find the following simple expressions\footnote{$\Gamma_{ij}^{k}|_{\textrm{hol}} = \frac{\partial y_k}{\partial t_a}
\frac{\partial^2 t_a} {\partial y_i \partial y_j}$}
\begin{align}
\Gamma_{12}^{2}|_{\textrm{hol}} &= \Gamma_{21}^{2}|_{\textrm{hol}} = \Gamma_{11}^{2}|_{\textrm{hol}}=0\quad\textrm{and}\quad \Gamma_{22}^{2}|_{\textrm{hol}} = - \frac{1}{y_2}\;.
\end{align}
The data we choose for constructing the polynomial building blocks is\footnote{$s^{k}_{ij}:=s^{y_k}_{y_i y_j}$.}
\begin{align}
s_{22}^{2} = - \frac{1}{y_2}\,,\quad s_{11}^{1} = -\frac{3}{2 y_1}\,, \quad s_{12}^1=s_{21}^{1} = - \frac{1}{4y_2}\,,
\end{align}
and all the other $s^k_{ij}$ zero. With this data only the propagator $S^{11}$ survives in the holomorphic limit but we know from the preceding discussion that also non-holomorphically there exists just one propagator with two indices.
We further find for the data of Eqs.~(\ref{finite9}-\ref{finite12})
\begin{equation}
 h_1^{11}=-\frac{y_1}{4}\,
\end{equation}
out of all the other holomorphic data only $h_2^{11}$is not zero in this setup but is not needed for the recursion. Now we still need the initial correlation functions to start the recursion. We only need one Yukawa coupling, namely
\begin{equation}
C_{111}= - (y_1^3\Delta)^{-1}\,, \quad \textrm{where} \quad \Delta= 1-8 y_1 (1+y_2)+ 16 y_1^2 (1-y_2)^2.
\end{equation}
And we further need  $\mathcal{F}_1^{(1)}$ to completely determine all the polynomials
\begin{equation}
\mathcal{F}^{(1)}_1= \frac{1}{6 y_1}-\frac{1}{2} \frac{S^{11}}{y_1^3 \Delta}-\frac{1}{12 \Delta} \left(32 y_1 (1-y_2)^2-8
   (1+y_2)\right).
\end{equation}
To obtain a bound on the maximal powers of $y_1$ and $y_2$ that we have to allow in the ansatz of the ambiguity we move on to region II.
\subsubsection*{\it Region II}
This is the orbifold expansion point in\cite{AKMV}. It is an orbifold in the moduli space of complex structures but does not correspond to  $\mathbbm{C}^3/\mathbbm{Z}_n$ target space on the A-model side. We find the right invariant coordinates in this patch to be $x_1=\frac{a_0}{\sqrt{a_1 a_2}}=\frac{1}{\sqrt{y_1}}$ and $x_2=\frac{a_1 a_2}{a_3 a_4}=y_2$. The mirror maps are given by
\begin{eqnarray}
t_1(x_1,x_2)&=& x_1 +\frac{1}{4} x_1 x_2 +\left(\frac{x_1^3}{24}+\frac{9}{64} x_2^2
   x_1\right) +\dots\,,
    \nonumber\\
t_2(x_2) &=& \log x_2\,.
\end{eqnarray}
We find that the propagator $S^{11}(t_1,t_2)$ is not regular on its own. We check however that the whole polynomial part of the amplitudes is regular. Here we show this for $\mathcal{F}^{(2)}$ which is expressed in terms of $t_1$ and $q_2=e^{t_2}$
\begin{equation}
\mathcal{F}^{(2)}(t_1,t_2)=\frac{1}{360}+\frac{1}{480} q_2+\left(\frac{31
   q_{2}^2}{3840}+\frac{t_1^2}{1920}\right)+\dots\,.
\end{equation}
Now we have enough information to make an ansatz for the ambiguity in terms of large complex structure coordinates. The discussion is more transparent in terms of the right coordinates at large complex structure $z_1$ and $z_2$. The problem is obviously symmetric in these coordinates. From region II we learn that we can make an ansatz which has as its highest degree monomials of the form $z_1^iz_2^{(n-i)}$ where $n$ refers to the highest degree of $z_1$ in the denominator in order to ensure regularity of the ansatz in $x_1$. In region III we get the same statement with the degree of $z_2$ in the denominator. So we can make a symmetric ansatz in the coordinates $z_1$ and $z_2$ of maximal joint degree $2 (2g-2)$. To fix the coefficients of the ansatz we have to use the gap condition at the conifold locus.
\subsubsection*{\it Conifold}
To parameterize the expansion around the conifold locus we pick the coordinates:
\begin{equation}
 u_1= \Delta\,, \quad \textrm{and} \quad u_2=1+y_2\,,
\end{equation}
the choice of the second coordinate is arbitrary, we only have to make sure that the coordinate is transverse to the discriminant. We solve the Picard-Fuchs equations and find the mirror maps
\begin{eqnarray}
 t_{1}(u_1,u_2) &=& u_1 +\frac{5}{8} u_1^2 +\left(\frac{89 u_1^3}{192}+\frac{3}{32} u_2^2
   u_1\right) +\dots \,,\nonumber\\
t_{2}(u_2) &=& \log(1+u_2)\,.
\end{eqnarray}
The propagator in the conifold coordinates now reads
\begin{equation}
\frac{S^{y_1 y_1}}{y_1^2}(t_1,t_2)= \frac{1}{4} t_1 -\frac{1}{8} t_1^2 +\frac{1}{768} \left(43
   t_1^3-6 t_1 t_2^2\right) +\dots \,.
\end{equation}
A counting of independent conditions for the free parameters in the ambiguity ansatz is more involved in this case. One would think that there are infinitely many conditions as we can move in the $u_2$ direction. This fact has also another manifestation. Requiring the vanishing of all but the leading singularity in the conifold coordinate involves setting series in the other coordinates to zero. It turns out that the conditions are not unrelated. Once the correct normalization of the mirror map at the conifold is chosen we find that the gap conditions with the contribution from constant maps are enough to fix the ambiguity up to genus $4$. We assume but cannot prove rigorously that this holds up to arbitrary genus.\footnote{This was also found in \cite{HKR}.}
\subsection{Local $\mathbbm{F}_2$}
Local $\mathbbm{F}_2$ denotes the anti-canonical bundle over $\mathbbm{F}_2$, which is obtained from
\begin{eqnarray}
l^{(a)}=\left(
\begin{array}{c|ccccccc}
0&0&0&-2&1&1&0&0\\
0&0&0&0&0&-2&1&1\\
-6&3&2&1&0&0&0&0
\end{array}\right)\,.
\end{eqnarray}
by decompactification. Much of the discussion here will follow the last example. The finite dual period in this case is
\begin{equation}
 (\p_{t_1}-\frac{1}{2} \p_{t_3}) \mathcal{F}=-\frac{1}{2}t_1 t_2-\frac{1}{2} t_1^2 +\dots\,.  
\end{equation}
After the change of coordinates
\begin{equation}
 t=t_1\,, \quad u=t_2\,, \quad s=t_3+ \frac{1}{2} t_1\,,
\end{equation}
the dual period becomes
\begin{equation}
\p_t \mathcal{F}= -\frac{1}{2}t(u+t)+ \dots
\end{equation}
Integrating this we obtain for the prepotential
\begin{equation}
 \mathcal{F}=-\frac{1}{6} t^3 -\frac{1}{4} t^2 u+\frac{a}{6} u^3 +\dots\,,
\end{equation}
with $a$ the arbitrary constant that we will set to zero. The nonzero ``classical intersections'' are the same as in the last model when we changed the K\"ahler parameters
\begin{equation} \label{intersectionf2}
 C^{(0)}_{ttt}=-1\,, \quad C^{(0)}_{sss}=8\,,  \quad C_{uss}^{(0)}=2\,, \quad \textrm{and} \quad C^{(0)}_{ttu}=-\frac{1}{2}\,.
\end{equation}
The analysis of taking $s$ to infinity also gives the same result and again there exists only one non-trivial propagator $S^{tt}$ and one holomorphic anomaly equation for the non-compact model.\\
\begin{figure}[t!]
\centering
\begin{minipage}{0.7\textwidth}
\centering
\subfigure[ A-model.]{
\centering
\psset{unit=1.7cm}
\begin{pspicture}(-1,-2)(1,1)
\psline[linewidth=1pt](0,0)(1,0)
\psline[linewidth=1pt](0,0)(0,1)
\psline[linewidth=1pt](0,0)(-1,-2)
\psline[linewidth=1pt](0,0)(0,-1)
\psline[linewidth=1pt](1,0)(0,1)
\psline[linewidth=1pt](0,1)(-1,-2)
\psline[linewidth=1pt](-1,-2)(0,-1)
\psline[linewidth=1pt](0,-1)(1,0)

\rput[tl](0.07,-1.03){$a_2$}
\rput[bl](0.07,1.03){$a_1$}
\rput[tr](-1.07,-2.03){$a_4$}
\rput[bl](1.03,0.07){$a_3$}
\rput[bl](0.07,0.07){$a_0$}

\psdot[dotsize=5pt 1,dotstyle=o](-1,-2)%
\psdot[dotsize=5pt 1,dotstyle=o](1,0)%
\psdot[dotsize=5pt 1,dotstyle=o](0,-1)
\psdot[dotsize=5pt 1,dotstyle=o](0,1)%
\psset{fillcolor=black}
\psdot[dotsize=5pt 1,dotstyle=o](0,0)%
\end{pspicture}
}\hfill
\subfigure[ B-model.]{
\centering
\raisebox{0.4cm}[0cm]{
\psset{unit=1.4cm}
\psset{griddots=0,gridlabels=0,subgriddiv=5}
\begin{pspicture}(-2,-2)(1,1)
\psgrid[griddots=5,subgriddiv=0,gridcolor=gray]

\psline[arrows=->,arrowsize=5pt,arrowinset=0](0,0)(1,0)
\psline[arrows=->,arrowsize=5pt,arrowinset=0](0,0)(0,1)
\psline[arrows=->,arrowsize=5pt,arrowinset=0](0,0)(1,-2)
\psline[arrows=->,arrowsize=5pt,arrowinset=0](0,0)(-2,0)

\rput[l](1.07,0){$(1,0)$}
\rput[b](0,1.07){$(0,1)$}
\rput[r](-2.07,0){$(-2,0)$}
\rput[t](1.07,-2.07){$(1,-2)$}
\rput[bl](-1.8,0.09){$a_0$}
\rput[br](0.8,0.09){$a_1$}
\rput[tl](0.07,0.8){$a_3,a_4$}
\rput[bl](0.9,-1.75){$a_2$}
\rput[tr](0.9,0.9){I}
\rput[br](0.9,-0.9){II}
\rput[bl](-0.9,-0.9){III}
\rput[tl](-0.9,0.9){IV}
\end{pspicture}}}
\caption{Fan and secondary fan for local $\mathbbm{F}_2$.}
\label{fig_f2}
\end{minipage}
\end{figure}
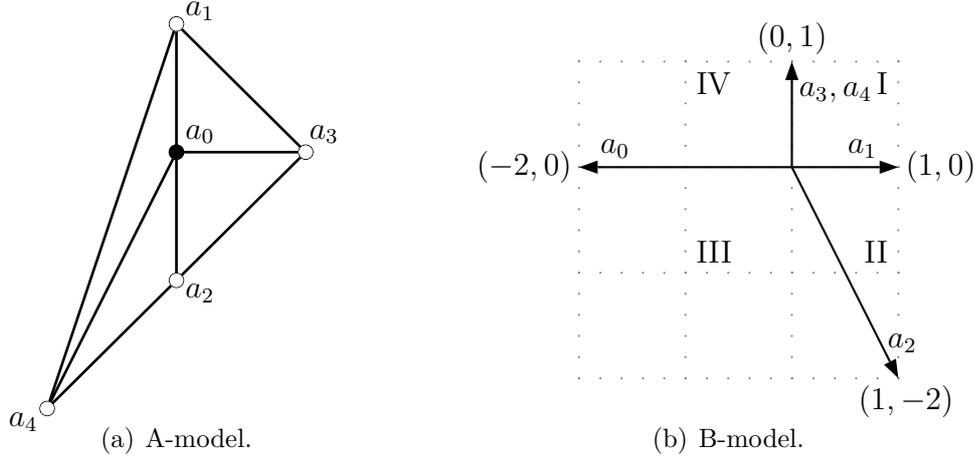
Local $\mathbbm{F}_2$ is described by the charge vectors
\begin{eqnarray}
l^{(a)}=\left(
\begin{array}{ccccc}
a_0&a_1&a_2&a_3&a_4\\
-2&1&1&0&0\\
0&0&-2&1&1
\end{array}\right)\, ,
\end{eqnarray}
which represents a $\mathbbm{P}^1$ fibered over $\mathbbm{P}^1$. $t_1$ will denote the K\"ahler parameter corresponding to the fiber and $t_2$ coresponds to the base.
The mirror geometry is given by,
\begin{align}
P &= a_0 u_1 u_2 u_3 + a_1 u_1^2 + a_2 u_2^2 u_3^2 + a_3 u_3^4 +a_4
u_2^4 =0\,.
\end{align}
The rescaling $u_i \rightarrow \lambda_i u_i$, $\lambda_i \in
\mathbbm{C}^*$ induces a $({\mathbbm{C}^*})^3$-action
\begin{align}
(a_0,a_1,a_2,a_3,a_4) &\rightarrow (\lambda_1 \lambda_2 \lambda_3
a_0,\lambda_1^2 a_1,\lambda_2^2 \lambda_3^2 a_2,\lambda_3^4
a_3, \lambda_2^4 a_4)\,.
\end{align}

The Picard-Fuchs operators are
\begin{align}
\mathcal{L}_1 & = \theta_{a_1} \theta_{a_2}  - \frac{a_1 a_2} {a_0^2} \theta_{a_0}(\theta_{a_0} - 1)\,, \\
\mathcal{L}_2 &= \theta_{a_3}  \theta_{a_4}  - \frac{a_3 a_4} {a_2^2} \theta_{a_2}(\theta_{a_2}-1)\,.
\end{align}

\subsubsection*{\it  Large Complex Structure, Region I}
In this region of the secondary fan we use the ${\mathbbm{C}^*}^3$-action to set
$a_0,a_2,a_4 \rightarrow 1$ and $a_1 \rightarrow a_0^{-2} a_1
a_2$, $a_3 \rightarrow a_2^{-2} a_3 a_4$ (alternatively $a_0,a_2,a_3
\rightarrow 1$ and $a_1 \rightarrow a_0^{-2} a_1
a_2$, $a_4 \rightarrow a_2^{-2} a_3 a_4$). In
both cases the good ${\mathbbm{C}^*}^3$-invariant
coordinates are given by
\begin{equation}
z_1 = \frac{a_1 a_2}{a_0^2}\, ,\quad z_2 = \frac{a_3 a_4} {a_2^2}\,.
\end{equation}
We find for the mirror maps
\begin{eqnarray}
t_1(z_1,z_2)&=& \log (z_1)+(2 z_1-z_2)+\left(3 z_1^2-\frac{3 z_2^2}{2}\right)\, 
   +\left(\frac{20 z_1^3}{3}+6 z_2 z_1^2-\frac{10 z_2^3}{3}\right)
  + \dots \nonumber\\
t_2(z_2)&=&\log (z_2)+2 z_2 +3 z_2^2 +\frac{20}{3} z_2^3 + \dots \, .
\end{eqnarray}
As the second mirror map only depends on $z_2$, the holomorphic limit of Christoffel connections with upper index $2$ and mixed lower indices vanish. We choose as holomorphic data
\begin{eqnarray}
 s_{11}^1 =-\frac{3}{2 z_1}, \quad s_{12}^1=s_{21}^1=-\frac{1}{4 z_2}\, .
\end{eqnarray}
It is also possible to choose $s_{22}^2$ and $s_{22}^1$ such that only the propagator $S^{11}$ survives. This is due to the special form of the mirror map. We will not need these for the recursion. All other $s_{ij}^k$ are set to zero. We find for the other holomorphic quantities
\begin{eqnarray}
 h_{1}^{11}= -\frac{z_1}{4},
\end{eqnarray}
also $h_{2}^{11}$ is nonzero but again not needed in the following. All the other holomorphic quantities are zero. The only Yukawa coupling relevant for the setup is given by
\begin{eqnarray}
C_{111}&=&-(z_1^3\Delta_1)^{-1}, \quad \Delta_1=(1-4 z_1)^2-64 z_1^2 z_2.
\end{eqnarray}
There are however two discriminants, the second is given by $\Delta_2=1-4z_2$. We will exmine the locus where $\Delta_2=0$ later on.
\begin{equation}
\mathcal{F}^{1}_1= \frac{1}{6z_1}-\frac{S^{11}}{2\Delta_1 z_1^3}  +\frac{8}{\Delta_1}  \left((1-4 z_1)+16 z_1 z_2 \right).
\end{equation}

{}From arguments in sect. 2.3. and equation \eqref{Gapp} it follows that 
the holomorphic ambiguity has the form $$\frac{f^{(g)}(z_1,z_2)}{\Delta_1^{2g-2} \Delta_2^{g-1}},$$
see below for a careful study of the local expansions at the singular loci. 
Furthermore the regularity of the polynomial part of the amplitudes at the orbifold expansion point in region III allows us again to put some bounds on the monomials that appear in $f^{g}(z_1,z_2)$. We find that the monomial of maximal degree at genus $g$ has the form $z_1^{4(g-1)}z_2^{3(g-1)}$ moreover every monomial $z_1^n z_2^m$ has to satisfy $m \le \frac{n}{2}+g-1$.

\subsubsection*{\it Orbifold, Region III}
In region III (orbifold region) of the secondary fan, i.e.~$a_1,a_3,a_4
\not= 0$
we use the ${\mathbbm{C}^*}^3$-action to set
$a_1,a_3,a_4 \rightarrow 1$ and $a_0 \rightarrow a_0 a_1^{-1/2}
(a_3a_4)^{-1/4}$,
$a_2 \rightarrow a_2 (a_3a_4)^{-1/2}$. Therefore the
good ${\mathbbm{C}^*}^3$-invariant coordinates are given by
\begin{equation}
x_1 = \frac{a_0}{a_1^{1/2} (a_3a_4)^{1/4}} = \frac{1}{z_1^{1/2} z_2^{1/4}}\,, \quad x_2 = \frac{a_2}{(a_3a_4)^{1/2}} = \frac{1}{z_2^{1/2}}\,.
\end{equation}

Under monodromy at large complex structure $z_1 \rightarrow \exp(2 \pi \text{i})z_1 $ we have $x_1
\rightarrow \alpha x_1$ and $x_2 \rightarrow x_2$ with $\alpha = \exp( -2
\pi \text{i} /2 )$. Under $z_2 \rightarrow \exp(2 \pi \text{i})z_2$ we
have $x_1 \rightarrow \beta x_1$ and $x_2 \rightarrow \beta^2 x_2$ with
$\beta = \exp( -2 \pi \text{i} /4 )$.
We fix the basis of solutions to the Picard-Fuchs equations
\begin{align}
w_0 &= 1\,, \\
w_1 &= x_1  + \frac{1}{32} x_1 x_2^2 -
\frac{1}{192} x_1^3 x_2 + \frac{1}{2560} x_1^5 + \frac{25}{6144} x_1
x_2^4 + \cdots\,, \\
w_2 &= x_2 + \frac{1}{24} x_2^3 + \frac{3}{640} x_2^5 + \cdots\,, \\
w_3 &= x_1 x_2  - \frac{1}{12} x_1^3  + \frac{3}{32} x_1 x_2^3 -
\frac{3}{128} x_1^3 x_2^2 + \cdots\,,
\end{align}
such that they transform under the monodromy as
\begin{align}
w_0 &\rightarrow w_0\,, \\
w_1 &\rightarrow \alpha \beta w_1\,, \\
w_2 &\rightarrow \beta^2 w_2\,, \\
w_3 &\rightarrow \alpha \beta^3 w_3\,.
\end{align}
We want to write this set of solutions in the form
$(1,t_1,t_2,\partial_{t_1} \mathcal{F})$ such that $\mathcal{F}$ is
monodromy invariant. The only possibility is given by $t_1 \propto w_1$,
$t_2 \propto w_2$ and $\partial_{t_1} \mathcal{F} \propto w_3$.
We normalize the mirror maps as
\begin{equation}
t_1 = w_1\,, \quad t_2 = w_2\,,
\end{equation}
with inverse
\begin{align}
x_1 &= t_1  - \frac{1}{32} t_1 t_2^2 +
\frac{1}{192} t_1^3 t_2 - \frac{1}{2560} t_1^5 - \frac{1}{2048} t_1
t_2^4 + \cdots\,. \\
x_2 &= t_2 - \frac{1}{24} t_2^3 + \frac{1}{1920} t_2^5 + \cdots\,,
\end{align}
and we normalize the prepotential as
\begin{align}
\partial_{t_1} \mathcal{F} &= \frac{1}{4} w_3 \;,
\end{align}
to obtain the correct Orbifold Gromov-Witten invariants at genus zero which were computed in \cite{Coates,Brini}. We read off the orbifold Gromov-Witten invariants at genus $g$ from the expansion of the topological string amplitudes on the A-model side in the two mirror maps $t_1$ and $t_2$ according to the formula 
$$
\mathcal{F}^{(g)} = \sum_{n_1 n_2=0}^\infty \frac{1}{n_1! n_2!}
N^{\text{orb}}_{g,(n_1 , n_2)} t_1^{n_1} t_2^{n_2},
$$
where $N^{\text{orb}}_{g,(n_1 , n_2)}$ denote the orbifold Gromov-Witten invariants. The results agree for genus 0 and 1 with \cite{Brini,Coates}. The higher genus invariants are new results of our analysis. We list these invariants up to genus 4 in the appendix. 

\subsubsection*{\it Conifold, $\Delta_1=0$}
We take coordinates
\begin{equation}
 y_1=\Delta_1 \quad \textrm{and} \quad y_2=z_2-1
\end{equation}
and find the mirror maps
\begin{eqnarray}
t_1(y_1,y_2)&=& y_1 +\left(\frac{87 y_1^2}{128}-\frac{1}{4} y_1 y_2\right)+\frac{\left(26217 y_1^3-8736 y_2 y_1^2+9728 y_2^2 y_1\right) }{49152}+\dots \\\nonumber
t_2(y_2)&=&y_2 -\frac{5}{6} y_2^2 +\frac{7}{9} y_2^3 + \dots\, .
\end{eqnarray}
As we can again move in the $y_2$ direction we could implement the gap condition at infinitely many different points. In this model we furthermore have a second discriminant. We will examine the behavior of the amplitudes at the locus where that discriminant vanishes in the following.

\subsubsection*{\it Singularity at $\Delta_2=0$}
We choose as coordinates 
\begin{equation}
 u_1=z_1, \quad u_2=\Delta_2
\end{equation}
and find the mirror maps
\begin{eqnarray}
t_1 &=& \log (u_1)+\left(2 u_1-\frac{u_2}{2}\right) +\left(\frac{9 u_1^2}{2}-\frac{u_2^2}{4}\right)+\frac{1}{6} \left(100 u_1^3-9 u_2 u_1^2-u_2^3\right) +\dots\nonumber\\
t_2 &=&\sqrt{u_2} +\frac{1}{3} u_2^{3/2} +\frac{1}{5} u_2^{5/2} +\dots \,.
\end{eqnarray}
corresponding to $\gamma=1/2$ in \eqref{Gapp}.
At the singular locus $\Delta_2=0$ the space-time spectrum contains extra massless states from an enhanced $SU(2)$ gauge symmetry
with one adjoint hypermultiplet \cite{scs}. According to  eq.\eqref{Gapp} the theory is regular due to the 
cancellation of the effects of the 
equal number of extra massless hyper and vector multiplets, $b=0$. We find that the expansion of the polynomial part of the amplitudes is 
already regular in $t_2$; this allows us to further restrict the ansatz of the holomorphic ambiguity to be of the form
$$\frac{f^{(g)}(z_1,z_2)}{\Delta_1^{2g-2}}.$$
Regularity at the orbifold expansion point requires the monomials to be of the type $z_1^nz_2^m$ with $m\le\frac{n}{2}$ and maximal degree of $n=2(2g-2)$.
\\
Once we have normalized the mirror map at the conifold locus to obtain the prefactor of Eq.~(\ref{Gap}) we find that the conditions obtained from implementing the gap are enough to supplement the polynomial part of the amplitudes with the correct holomorphic ambiguities. 
We refer to the appendix for the results. 
Since the gap conditions hold at infinitely many points, it is plausible
that these boundary conditions might be sufficient for arbitrary genus. 
This finishes our analysis of the examples of the polynomial construction on local models.

\section{Conclusion}
In this paper we studied the polynomial construction of \cite{Yau,AL} in the case of local Calabi-Yau manifolds. The polynomials provide expressions for the topological string amplitudes everywhere in moduli space. This can be used to study the topological string at different expansion points. In particular we analyzed the freedom in choosing the polynomial building blocks and how to exploit this freedom elaborately to maximize the information that can be obtained from various expansion points. We further clarified which simplifications can occur in the formalism on the computational side when local models are studied. As examples we studied local $\mathbbm{P}^2$, local $\mathbbm{F}_0$ and local $\mathbbm{F}_2$. An immediate application of the construction is the possibility to extract Orbifold Gromov-Witten invariants which have been already computed for $\mathbbm{C}^3/\mathbbm{Z}_3$ and to make predictions for higher genus Orbifold GW invariants for $\mathbbm{C}^3/\mathbbm{Z}_4$.
 
A simple counting for local $\mathbbm{P}^2$ shows that the information coming from the boundary conditions is enough to supplement the polynomials with the right holomorphic ambiguity at every genus. For the other two models we argue that the information coming from boundary conditions is enough for all genera 
but we cannot demonstrate this rigorously. However it appears that the boundary conditions at the various boundary divisors and  
divisors with extra massless states should provide enough information in general. Further information can be obtained by 
studying intersections of the singular divisors which are often described also by eq.\eqref{Gap}, 
with extra  massless states at the intersection point.

Having a powerful alternative computation method could help pushing forward the understanding in some directions which have been explored recently. The polynomial construction of topological string amplitudes 
applies also to compactifications with background D-branes \cite{Walcher,AL,Konishi}. 
It should be straightfoward to work out the corresponding boundary conditions also for this case and 
fix the holomorphic ambiguity along the lines of the above arguments.
\\
\\
{\bf Acknowledgments}\\
We would like to thank R.~Cavalieri, M.~Hecht, C.~R\"omelsberger, I.~Sachs and E.~Scheidegger
for helpful discussions. The work of M.A.and P.M.~is supported by the German Excellence Initiative via the program ``Origin and Structure of the Universe", and the work of  J.D.L. is supported by the DFG.
\appendix
\section{The Ambiguities} 
\label{apppol}
Using the method described in this work we obtained polynomial
expression for the topological string partition functions.
In this appendix we give the explicit expressions of some of
the holomorphic ambiguities fixed by the method discussed in the main
part of this paper.
\subsection{Local $\mathbbm{P}^2$}
{\scriptsize
\begin{eqnarray}
f^{(2)} &=& \frac{-216 P^2+4 P+1}{17280}\,,\nonumber \\ 
f^{(3)}&=&\frac{P^4}{112}-\frac{29 P^3}{3360}+\frac{2263 P^2}{1088640}-\frac{13
   P}{136080}-\frac{1}{4354560} \,,\nonumber \\ 
f^{(4)}&=&-\frac{3}{160} P^6+\frac{7639 P^5}{201600}-\frac{32957
   P^4}{1209600}+\frac{1211911 P^3}{146966400}-\frac{252559
   P^2}{261273600}+\frac{3121 P}{97977600}-\frac{311}{2351462400}\,. \nonumber
\end{eqnarray}
}
\newpage
\subsection{Local ${\mathbbm{F}_0}$}
In the following all $f^{(g)}$ are multiplied by $\Delta^{2-2g}$ to give the ambiguity at genus $g$.\\

\textbf{Genus 2}
{\scriptsize
\begin{eqnarray*}
f^{(2)} &=& -\frac{1} {60} + \frac{121}{720} (1 + y_2) y_1+\frac{1} {180} (-75 - 338 y_2 - 75 y_2^2) y_1^2 +\frac{1}{15} (-7 + 71 y_2 + 71 y_2^2 - 7 y_2^3) y_1^3\\
&&+\frac{4}{45} (-1 + y_2)^2 (23 + 50 y_2 + 23 y_2^2) y_1^4\,.
\end{eqnarray*}}
\textbf{Genus 3}
{\scriptsize \begin{eqnarray*}
f^{(3)} &=& \frac{10037} {2903040} - \frac{115}{2268} (1 + y_2) y_1+ \frac{1}{22680} (1607 + 42269 y_2 + 1607 y_2^2) y_1^2\\
&&+ \frac{1}{5670} (16699 - 91239 y_2 - 91239 y_2^2 +
16699 y_2^3) y_1^3 \\
&&+ \frac{1}{5670} (-137695 + 142484 y_2 + 1218774 y_2^2
+ 142484 y_2^3 - 137695 y_2^4) y_1^4 \\
&&+
\frac{16}{2835} (14653 + 42763 y_2 - 147656 y_2^2 - 147656 y_2^3 +
42763 y_2^4 + 14653 y_2^5) y_1^5 \\
&& - \frac{32}{2835}
(11000 + 95349 y_2 - 29772 y_2^2 - 337474 y_2^3 -
29772 y_2^4 + 95349 y_2^5 + 11000 y_2^6) y_1^6\\
&& +
\frac{128}{2835} (-1 + y_2)^2 (803 + 28023 y_2 + 109414 y_2^2 +
109414 y_2^3 + 28023 y_2^4 + 803 y_2^5) y_1^7 \\
&& + \frac{64}{2835} (-1 + y_2)^4 (2833 + 22172 y_2 + 42150 y_2^2 + 22172 y_2^3
2833 y_2^4) y_1^8\,.
\end{eqnarray*}}
\textbf{Genus 4}
{\scriptsize
\begin{eqnarray*}
f^{(4)} &=& -\frac{934993}{696729600} +
\frac{1873}{82944} (1 + y_2) y_1 +
\frac{1}{21772800}(1805481 - 35643448 y_2 + 1805481 y_2^2) y_1^2\\
&&+ \frac{1}{10886400}(-55395131 + 227680355 y_2 + 227680355 y_2^2 -
55395131 y_2^3) y_1^3\\
&& + \frac{1}{226800}(12490827 + 343564 y_2 - 139937742 y_2^2 +
343564 y_2^3 + 12490827 y_2^4) y_1^4 \\
&& + \frac{1}{170100} (-48921165 - 310982873 y_2 +
864161766 y_2^2 + 864161766 y_2^3 - 310982873 y_2^4 -
48921165 y_2^5) y_1^5 \\
&& + \frac{2}{42525}  (14314083 + 327499585 y_2 - 71267327 y_2^2
- 1601136842 y_2^3 - 71267327 y_2^4 + 327499585 y_2^5 + 14314083 y_2^6)
y_1^6 \\
&& + \frac{2}{14175}  (3603345 - 394817549 y_2 - 1102751935 y_2^2 +
2494798139 y_2^3 + 2494798139 y_2^4 - 1102751935 y_2^5 \\
&& - 394817549 y_2^6 + 3603345 y_2^7) y_1^7 \\
&& - \frac{4}{42525} (82804869 - 778119160 y_2 - 8753666580 y_2^2 +
1161603000 y_2^3 + 26815280030 y_2^4 + 1161603000 y_2^5 \\
&& - 8753666580 y_2^6 - 778119160 y_2^7 + 82804869 y_2^8) y_1^8 \\
&& + \frac{128}{42525} (7059047 + 32956639 y_2 - 493715972 y_2^2 -
1075054388 y_2^3 + 1776480754 y_2^4 + 1776480754 y_2^5 -
1075054388 y_2^6 \\
&& - 493715972 y_2^7 + 32956639 y_2^8 +
7059047 y_2^9) y_1^9 \\
&& - \frac{128}{14175} (-1 + y_2)^2 (2818187 + 50363322 y_2 +
78760776 y_2^2 - 644284250 y_2^3 - 1452576870 y_2^4 - 644284250 y_2^5\\
&& + 78760776 y_2^6 + 50363322 y_2^7 + 2818187 y_2^8) y_1^{10} \\
&& + \frac{256}{4725} (-1 + y_2)^4 (168629 + 6546911 y_2 + 54108709
y_2^2 + 145614151 y_2^3 + 145614151 y_2^4 + 54108709 y_2^5 \\
&& + 6546911 y_2^6 + 168629 y_2^7) y_1^{11} \\
&& + \frac{2048}{42525} (-1 + y_2)^6 (85909 + 1579674 y_2
+ 7561563 y_2^2 + 12511468 y_2^3 + 7561563 y_2^4 + 1579674 y_2^5 +
85909 y_2^6) y_1^{12}\,.
\end{eqnarray*}}
\subsection{Local ${\mathbbm{F}_2}$}
In the following all $f^{(g)}$ are multiplied by $\Delta_1^{2-2g}$ to give the ambiguity at genus $g$. \\

\textbf{Genus 2}
{\scriptsize
\begin{eqnarray*}
f^{(2)}&=&-\frac{1}{60}+\frac{121}{720} z_1 -\frac{5}{12} z_1^2 -\frac{7}{15}
   z_1^3-\frac{47}{45} z_2 z_1^2 +\frac{92 z_1^4}{45}+\frac{92}{15}
   z_2 z_1^3 -\frac{352}{45} z_1^4 z_2 -\frac{64}{45}z_1^4 z_2^2\, .
\end{eqnarray*}}
\textbf{Genus 3}
{\scriptsize
\begin{eqnarray*}
f^{(3)}&=&
\frac{10037}{2903040}-\frac{115 z_1}{2268}+\frac{1607 z_1^2
   }{22680}+\frac{16699 z_1^3}{5670}+\frac{7811 z_2 z_1^2}{4536}
   -\frac{27539 z_1^4}{1134}-\frac{7852}{315} z_2 z_1^3
   +\frac{234448 z_1^5}{2835}+\\&& \frac{115544}{945} z_2 z_1^4 
 -\frac{70400}{567} z_1^6-\frac{488032 z_2
   z_1^5}{2835}+\frac{109736}{945} z_2^2 z_1^4 +\frac{102784
   z_1^7}{2835}-\frac{34784}{105} z_2 z_1^6-\frac{72064}{63} z_2^2 z_1^5
  \\ &&+\frac{181312 z_1^8}{2835}+\frac{887296}{945} z_2
   z_1^7 
+\frac{3329792}{945} z_2^2 z_1^6 -\frac{756736 z_2
   z_1^8}{2835}-\frac{948224}{315} z_2^2 z_1^7+\frac{3495424 z_2^3
   z_1^6}{2835} -\frac{161792}{189} z_2^2 z_1^8\\&& -\frac{3006464}{567}
   z_2^3 z_1^7 +\frac{9322496 z_1^8 z_2^3}{2835}+\frac{507904}{405} z_1^8 z_2^4 \, .
\end{eqnarray*}}
\textbf{Genus 4}
{\scriptsize
\begin{eqnarray*}
f^{(4)}&=&
-\frac{934993}{696729600}+\frac{1873 z_1 }{82944}+\frac{200609 z_1^2}{2419200}-\frac{55395131 z_1^3}{10886400}-\frac{3925441 z_2
   z_1^2}{2177280} +\frac{4163609 z_1^4}{75600}+\frac{98466437 z_2
   z_1^3}{2721600} \\ 
&&-\frac{120793}{420} z_1^5-\frac{3101234 z_2
   z_1^4}{14175} +\frac{1363246 z_1^6}{2025}-\frac{16594262 z_2
   z_1^5}{42525}-\frac{7227701 z_2^2 z_1^4}{14175} +\frac{480446
   z_1^7}{945}+\frac{69032882 z_2 z_1^6}{6075}\\ 
&&+\frac{25875076 z_2^2
   z_1^5}{2835} -\frac{1752484}{225} z_1^8-\frac{120011704 z_2
   z_1^7}{2025} -\frac{55663952}{945} z_2^2 z_1^6 +\frac{903558016
   z_1^9}{42525}+\frac{823176064 z_2 z_1^8}{6075}\\
  &&+\frac{122904352}{945} z_2^2
   z_1^7-\frac{1664462368 z_2^3 z_1^6}{42525} -\frac{360727936
   z_1^{10}}{14175}-\frac{3913572352 z_2 z_1^9}{42525}+\frac{647694464 z_2^2
   z_1^8}{2835}+\frac{7607485568 z_2^3 z_1^7}{14175}\\ 
&&+\frac{43169024 z_1^{11}}{4725}-\frac{235307776 z_2
   z_1^{10}}{1575}-\frac{22776242176 z_2^2 z_1^9}{14175}-\frac{12377030656 z_2^3
   z_1^8}{4725} +\frac{175941632 z_1^{12}}{42525}+\frac{1028473856
   z_2 z_1^{11}}{4725}\\ 
&&+\frac{1017880576}{405} z_2^2 z_1^{10}+\frac{210322817024
   z_2^3 z_1^9}{42525}-\frac{34826355712 z_2^4 z_1^8}{42525}
   +\frac{22740992 z_2 z_1^{12}}{14175}-\frac{281583616}{315} z_2^2
   z_1^{11}\\&& -\frac{14985371648 z_2^3 z_1^{10}}{14175}+\frac{302835073024 z_2^4
   z_1^9}{42525} 
 -\frac{129335296}{405} z_2^2
   z_1^{12}-\frac{22828384256 z_2^3 z_1^{11}}{4725}-\frac{89728712704 z_2^4
   z_1^{10}}{4725} \\
&&+\frac{8916041728 z_2^3
   z_1^{12}}{8505}+\frac{66821619712 z_2^4 z_1^{11}}{4725}-\frac{52007469056 z_2^5
   z_1^{10}}{14175} +\frac{3297247232 z_2^4
   z_1^{12}}{2835}+\frac{20264517632 z_2^5 z_1^{11}}{1575}
   \\&& -\frac{75765907456 z_1^{12} z_2^5
   }{14175}-\frac{7038042112 z_1^{12} z_2^6}{6075} \, .
\end{eqnarray*}}
\newpage
\section{Gopakumar-Vafa and Orbifold Gromov-Witten Invariants}
Replacing the generators by their holomorphic limits we can extract
the Gopakumar-Vafa invariants from the partition functions.
\label{tables}
\subsection{Local $\mathbbm{P}^2$}
\textbf{Orbifold Gromov-Witten Invariants} 
 \renewcommand{\arraystretch}{1.5}
\begin{center}
{\scriptsize
\begin{tabular*}{0.9\textwidth}{@{\extracolsep{\fill}} | c | c c c c c | }
\hline
$g \backslash d $&$ 0 $&$ 1 $&$ 2 $&$ 3 $&$ 4$ \\ \hline
$ 0$&$0 $&$ \frac{1}{3} $&$ -\frac{1}{27} $&$ \frac{1}{9} $&$ -\frac{1093}{729}$ \\
$ 1$&$0 $&$ 0 $&$ \frac{1}{243} $&$ -\frac{14}{243} $&$ \frac{13007}{6561}$ \\
$ 2$&$\frac{1}{17280} $&$ \frac{1}{19440} $&$ -\frac{13}{11664} $&$
\frac{20693}{524880} $&$
   -\frac{12803923}{4723920}$ \\
$ 3$&$-\frac{1}{4354560} $&$ -\frac{31}{2449440} $&$ \frac{11569}{22044960} $&$
   -\frac{2429003}{66134880} $&$ \frac{871749323}{198404640}$ \\
$ 4$&$-\frac{311}{2351462400} $&$ \frac{313}{62985600} $&$ -\frac{1889}{5038848} $&$
   \frac{115647179}{2550916800} $&$ -\frac{29321809247}{3401222400}$\\ \hline
\end{tabular*}}
\end{center}
  \renewcommand{\arraystretch}{1.0}
\subsection{Local ${\mathbbm{F}_0}$}
\textbf{Gopakumar-Vafa Invariants}\\ 
\textbf{ Genus 0}
\begin{center}
{\scriptsize
\begin{tabular*}{0.9\textwidth}{@{\extracolsep{\fill}} | c | c c c c c c c c | }
 \hline
$d_1 \backslash d_2$ & 0 & 1 & 2 & 3 & 4 & 5 & 6 & 7\\ \hline
0 & 0 & -2 & 0 & 0 & 0 & 0 & 0 & 0\\
1 & -2 & -4 & -6 & -8 & -10 & -12 & -14 & -16\\
2 & 0 & -6 & -32 & -110 & -288 & -644 & -1280 & -2340 \\
3 & 0 & -8 & -110 & -756 & -3556 & -13072 & -40338 & -109120\\
4 & 0 & -10 & -288 & -3556 & -27264 & -153324 & -690400 & -2627482 \\
5 & 0 & -12 & -644 & -13072 & -153324 & -1252040 & -7877210 & -40635264 \\
6 & 0 & -14 & -1280 & -40338 & -690400 & -7877210 & -67008672 &-455426686 \\
7 & 0 & -16 & -2340 & -109120 & -2627482 & -40635264 & -455426686 &
-3986927140 \\ \hline
\end{tabular*}}
\end{center}
\textbf{Genus 1}
\begin{center}
{\scriptsize
\begin{tabular*}{0.9\textwidth}{@{\extracolsep{\fill}} | c | c c c c c c c c | }
\hline
$d_1 \backslash d_2$ & 0 & 1 & 2 & 3 & 4 & 5 & 6 & 7\\ \hline
0 & 0 & 0 & 0 & 0 & 0 & 0 & 0 & 0 \\
1 & 0 & 0 & 0 & 0 & 0 & 0 & 0 & 0 \\
2 & 0 & 0 & 9 & 68 & 300 & 988 & 2698 & 6444 \\
3 & 0 & 0 & 68 & 1016 & 7792 & 41376 & 172124 & 599856 \\
4 & 0 & 0 & 300 & 7792 & 95313 & 760764 & 4552692 & 22056772 \\
5 & 0 & 0 & 988 & 41376 & 760764 & 8695048 & 71859628 & 467274816 \\
6 & 0 & 0 & 2698 & 172124 & 4552692 & 71859628 & 795165949 &6755756732
\\
7 & 0 & 0 & 6444 & 599856 & 22056772 & 467274816 & 6755756732 &
73400088512 \\ \hline
\end{tabular*}}
\end{center}
\textbf{Genus 2}
\begin{center}
{\scriptsize
\begin{tabular*}{0.9\textwidth}{@{\extracolsep{\fill}} | c | c c c c c c c c | }
\hline
$d_1 \backslash d_2$ & 0 & 1 & 2 & 3 & 4 & 5 & 6 & 7 \\ \hline
0 & 0 & 0 & 0 & 0 & 0 & 0 & 0 & 0 \\
1 & 0 & 0 & 0 & 0 & 0 & 0 & 0 & 0\\
2 & 0 & 0 & 0 & -12 & -116 & -628 & -2488 & -8036 \\
3 & 0 & 0 & -12 & -580 & -8042 & -64624 & -371980 & -1697704 \\
4 & 0 & 0 & -116 & -8042 & -167936 & -1964440 & -15913228 & -99308018 \\
5 & 0 & 0 & -628 & -64624 & -1964440 & -32242268 & -355307838 &
-2940850912 \\
6 & 0 & 0 & -2488 & -371980 & -15913228 & -355307838 & -5182075136 &
-55512436778 \\
7 & 0 & 0 & -8036 & -1697704 & -99308018 & -2940850912 & -55512436778 &
-754509553664 \\ \hline
\end{tabular*}}
\end{center}
\textbf{Genus 3}
\begin{center}
{\scriptsize
\begin{tabular*}{0.9\textwidth}{@{\extracolsep{\fill}} | c | c c c c c c c c | }
\hline
$d_1 \backslash d_2$ & 0 & 1 & 2 & 3 & 4 & 5 & 6 & 7\\ \hline
0 & 0 & 0 & 0 & 0 & 0 & 0 & 0 & 0 \\
1 & 0 & 0 & 0 & 0 & 0 & 0 & 0 & 0 \\
2 & 0 & 0 & 0 & 0 & 15 & 176 & 1130 & 5232 \\
3 & 0 & 0 & 0 & 156 & 4680 & 60840 & 501440 & 3059196 \\
4 & 0 & 0 & 15 & 4680 & 184056 & 3288688 & 36882969 & 300668468 \\
5 & 0 & 0 & 176 & 60840 & 3288688 & 80072160 & 1198255524 & 12771057936 \\
6 & 0 & 0 & 1130 & 501440 & 36882969 & 1198255524 & 23409326968 &
319493171724 \\
7 & 0 & 0 & 5232 & 3059196 & 300668468 & 12771057936 & 319493171724 &
5485514375644 \\ \hline
\end{tabular*}}
\end{center}
\textbf{Genus 4}
\begin{center}
{\scriptsize
\begin{tabular*}{0.9\textwidth}{@{\extracolsep{\fill}} | c | c c c c c c c c | }
\hline
$d_1 \backslash d_2$ & 0 & 1 & 2 & 3 & 4 & 5 & 6 & 7\\ \hline
0 & 0 & 0 & 0 & 0 & 0 & 0 & 0 & 0 \\
1 & 0 & 0 & 0 & 0 & 0 & 0 & 0 & 0 \\
2 & 0 & 0 & 0 & 0 & 0 & -18 & -248 & -1842 \\
3 & 0 & 0 & 0 & -16 & -1560 & -36408 & -450438 & -3772316 \\
4 & 0 & 0 & 0 & -1560 & -133464 & -3839632 & -61250176 & -662920988 \\
5 & 0 & 0 & -18 & -36408 & -3839632 & -144085372 & -2989287812 &
-41557026816 \\
6 & 0 & 0 & -248 & -450438 & -61250176 & -2989287812 & -79635105296 &
-1400518786592 \\
7 & 0 & 0 & -1842 & -3772316 & -662920988 & -41557026816 &
-1400518786592 & -30697119068800 \\ \hline
\end{tabular*}}
\end{center}

\subsection{Local ${\mathbbm{F}_2}$}

\subsubsection*{Gopakumar-Vafa Invariants}
$d_1$ and $d_2$ denote the degrees of the fiber and base classes respectively.\footnote{Note that
the correct genus zero data gives a value $n^{(0)}_{0,1}=0$ which is different from the 
naive result $n^{(0)}_{0,1}=-\frac{1}{2}$ obtained from local mirror symmetry in \cite{Chiang}.}\\
\textbf{Genus 0}
\begin{center}
{\scriptsize
\begin{tabular*}{0.9\textwidth}{@{\extracolsep{\fill}} | c | c c c c c c c c | }
\hline
$d_1 \backslash d_2$ & 0 & 1 & 2 & 3 & 4 & 5 & 6 & 7\\ \hline
0&0 & ${\bf 0}$ & 0 & 0 & 0 & 0 & 0 & 0\\
1& -2 & -2 & 0 & 0 & 0 & 0 & 0 & 0\\
2& 0 & -4 & 0 & 0 & 0 & 0 & 0 & 0 \\
3& 0 & -6 & -6 & 0 & 0 & 0 & 0 & 0 \\
4& 0 & -8 & -32 & -8 & 0 & 0 & 0 & 0 \\
5& 0 & -10 & -110 & -110 & -10 & 0 & 0 & 0 \\
6& 0 & -12 & -288 & -756 & -288 & -12 & 0 & 0 \\
7& 0 & -14 & -644 & -3556 & -3556 & -644 & -14 & 0 \\
8& 0 & -16 & -1280 & -13072 & -27264 & -13072 & -1280 & -16\\
\hline 
\end{tabular*}}
\end{center}
\textbf{Genus 1}
\begin{center}
{\scriptsize
\begin{tabular*}{0.9\textwidth}{@{\extracolsep{\fill}} | c | c c c c c c c c | }
\hline
$d_1 \backslash d_2$ & 0 & 1 & 2 & 3 & 4 & 5 & 6 & 7\\ \hline
 0&0 & 0 & 0 & 0 & 0 & 0 & 0 & 0 \\
 1&0 & 0 & 0 & 0 & 0 & 0 & 0 & 0 \\
 2&0 & 0 & 0 & 0 & 0 & 0 & 0 & 0 \\
 3&0 & 0 & 0 & 0 & 0 & 0 & 0 & 0 \\
 4&0 & 0 & 9 & 0 & 0 & 0 & 0 & 0 \\
 5&0 & 0 & 68 & 68 & 0 & 0 & 0 & 0 \\
 6&0 & 0 & 300 & 1016 & 300 & 0 & 0 & 0\\
 7&0 & 0 & 988 & 7792 & 7792 & 988 & 0 & 0\\
 8&0 & 0 & 2698 & 41376 & 95313 & 41376 & 2698 & 0 \\
\hline
\end{tabular*}}
\end{center}
\newpage
\textbf{Genus 2}
\begin{center}
{\scriptsize
\begin{tabular*}{0.9\textwidth}{@{\extracolsep{\fill}} | c | c c c c c c c c | }
\hline
$d_1 \backslash d_2$ & 0 & 1 & 2 & 3 & 4 & 5 & 6 & 7\\ \hline
0& 0& 0 & 0 & 0 & 0 & 0 & 0 & 0 \\
1& 0 & 0 & 0 & 0 & 0 & 0 & 0 & 0 \\
2& 0 & 0 & 0 & 0 & 0 & 0 & 0 & 0 \\
3& 0 & 0 & 0 & 0 & 0 & 0 & 0 & 0 \\
4& 0 & 0 & 0 & 0 & 0 & 0 & 0 & 0 \\
5& 0 & 0 & -12 & -12 & 0 & 0 & 0 & 0 \\
6& 0 & 0 & -116 & -580 & -116 & 0 & 0 & 0 \\
7& 0 & 0 & -628 & -8042 & -8042 & -628 & 0 & 0 \\
8& 0 & 0 & -2488 & -64624 & -167936 & -64624 & -2488 & 0 \\
\hline
\end{tabular*}}
\end{center}
\textbf{Genus 3}
\begin{center}
{\scriptsize
\begin{tabular*}{0.9\textwidth}{@{\extracolsep{\fill}} | c | c c c c c c c c | }
\hline
$d_1 \backslash d_2$ & 0 & 1 & 2 & 3 & 4 & 5 & 6 & 7\\ \hline
0& 0 & 0 & 0 & 0 & 0 & 0 & 0 & 0\\
1& 0 & 0 & 0 & 0 & 0 & 0 & 0 & 0\\
2& 0 & 0 & 0 & 0 & 0 & 0 & 0 & 0\\
3& 0 & 0 & 0 & 0 & 0 & 0 & 0 & 0\\
4& 0 & 0 & 0 & 0 & 0 & 0 & 0 & 0\\
5& 0 & 0 & 0 & 0 & 0 & 0 & 0 & 0\\
6& 0 & 0 & 15 & 156 & 15 & 0 & 0 & 0\\
7& 0 & 0 & 176 & 4680 & 4680 & 176 & 0 & 0\\
8& 0 & 0 & 1130 & 60840 & 184056 & 60840 & 1130 & 0 \\
\hline
\end{tabular*}}
\end{center}
\textbf{Genus 4}
\begin{center}
{\scriptsize
\begin{tabular*}{0.9\textwidth}{@{\extracolsep{\fill}} | c | c c c c c c c c | }
\hline
$d_1 \backslash d_2$ & 0 & 1 & 2 & 3 & 4 & 5 & 6 & 7\\ \hline
0& 0 & 0 & 0 & 0 & 0 & 0 & 0 & 0\\
1& 0 & 0 & 0 & 0 & 0 & 0 & 0 & 0\\
2& 0 & 0 & 0 & 0 & 0 & 0 & 0 & 0\\
3& 0 & 0 & 0 & 0 & 0 & 0 & 0 & 0\\
4& 0 & 0 & 0 & 0 & 0 & 0 & 0 & 0\\
5& 0 & 0 & 0 & 0 & 0 & 0 & 0 & 0\\
6& 0 & 0 & 0 & -16 & 0 & 0 & 0 & 0\\
7& 0 & 0 & -18 & -1560 & -1560 & -18 & 0 & 0\\
8& 0 & 0 & -248 & -36408 & -133464 & -36408 & -248 & 0\\
\hline
\end{tabular*}}
\end{center}
\subsubsection{Orbifold Gromov-Witten invariants for $\mathbbm{C}^3/\mathbbm{Z}_4$}
\textbf{ Genus 0}
 \renewcommand{\arraystretch}{1.15}
\begin{center}
{\scriptsize
\begin{tabular*}{0.9\textwidth}{@{\extracolsep{\fill}} | c | c c c c c | }
\hline
$n_2 \backslash n_1 $ & $ 2 $ & $ 4 $ & $ 6 $ & $ 8 $ & $ 10 $ \\ \hline
$0 $ & $ 0 $ & $ -\frac{1}{8} $ & $ 0 $ & $ -\frac{9}{64} $ & $ 0 $ \\
$1 $ & $ \frac{1}{4} $ & $ 0 $ & $ \frac{7}{128} $ & $ 0 $ & $ \frac{1083}{1024} $ \\
$2 $ & $ 0 $ & $ -\frac{1}{32} $ & $ 0 $ & $ -\frac{143}{512} $ & $ 0 $ \\
$3 $ & $ \frac{1}{32} $ & $ 0 $ & $ \frac{3}{32} $ & $ 0 $ & $ \frac{85383}{16384} $ \\
$4 $ & $ 0 $ & $ -\frac{11}{256} $ & $ 0 $ & $ -\frac{159}{128} $ & $  0 $ \\
$5 $ & $ \frac{1}{32} $ & $ 0 $ & $ \frac{47}{128} $ & $ 0 $ & $ \frac{360819}{8192} $ \\
$6 $ & $ 0 $ & $ -\frac{147}{1024} $ & $ 0 $ & $ -\frac{157221}{16384} $ & $ 0 $ \\
$7 $ & $ \frac{87}{1024} $ & $ 0 $ & $ \frac{20913}{8192} $ & $ 0 $ & $
\frac{73893099}{131072} $ \\
\hline
\end{tabular*}}
\end{center}
\renewcommand{\arraystretch}{1.0}
\textbf{ Genus 1}
\renewcommand{\arraystretch}{1.15}
\begin{center}
{\scriptsize
\begin{tabular*}{0.9\textwidth}{@{\extracolsep{\fill}} | c | c c c c c | }
\hline
$n_2 \backslash n_1 $ & $ 2 $ & $ 4 $ & $ 6 $ & $ 8 $ & $ 10 $ \\ \hline
$0$ & $ 0 $ & $ \frac{1}{128} $ & $ 0 $ & $ \frac{441}{4096} $ & $ 0 $ \\
$1$ & $ -\frac{1}{192} $ & $ 0 $ & $ -\frac{31}{1024} $ & $ 0 $ & $ -\frac{71291}{32768} $ \\
$2$ & $ 0 $ & $ \frac{35}{3072} $ & $ 0 $ & $ \frac{235}{512} $ & $ 0 $ \\
$3$ & $ -\frac{5}{768} $ & $ 0 $ & $ -\frac{485}{4096} $ & $ 0 $ & $ -\frac{2335165}{131072} $ \\
$4$ & $ 0 $ & $ \frac{485}{12288} $ & $ 0 $ & $ \frac{458295}{131072} $ & $ 0 $ \\
$5$ & $ -\frac{39}{2048} $ & $ 0 $ & $ -\frac{40603}{49152} $ & $ 0 $ & $ -\frac{58775443}{262144} $ \\
$6$ & $ 0 $ & $ \frac{2025}{8192} $ & $ 0 $ & $ \frac{10768885}{262144} $ & $ 0 $ \\
$7$ & $ -\frac{2555}{24576} $ & $ 0 $ & $ -\frac{293685}{32768} $ & $ 0 $ & $ -\frac{522517275}{131072}$\\
\hline
\end{tabular*}}
\end{center}
\renewcommand{\arraystretch}{1.0}
\textbf{ Genus 2}
\renewcommand{\arraystretch}{1.15}
\begin{center}
{\scriptsize
\begin{tabular*}{0.9\textwidth}{@{\extracolsep{\fill}} | c | c c c c c | }
\hline
$n_2 \backslash n_1 $ & $ 2 $ & $ 4 $ & $ 6 $ & $ 8 $ & $ 10 $ \\ \hline
$0$ & $ 0 $ & $ -\frac{61}{30720} $ & $ 0 $ & $ -\frac{9023}{81920} $ & $ 0 $ \\
$1$ & $ \frac{41}{46080} $ & $ 0 $ & $ \frac{6061}{245760} $ & $ 0 $ & $ \frac{36213661}{7864320} $ \\
$2$ & $ 0 $ & $ -\frac{647}{92160} $ & $ 0 $ & $ -\frac{1066027}{1310720} $ & $ 0 $ \\
$3$ & $ \frac{257}{92160} $ & $ 0 $ & $ \frac{168049}{983040} $ & $ 0 $ & $ \frac{887800477}{15728640} $ \\
$4$ & $ 0 $ & $ -\frac{65819}{1474560} $ & $ 0 $ & $ -\frac{18530321}{1966080} $ & $ 0 $ \\
$5$ & $ \frac{23227}{1474560} $ & $ 0 $ & $ \frac{43685551}{23592960} $ & $ 0 $ & $ \frac{62155559923}{62914560} $ \\
$6$ & $ 0 $ & $ -\frac{437953}{983040} $ & $ 0 $ & $ -\frac{9817250341}{62914560} $ & $ 0 $ \\
$7$ & $ \frac{418609}{2949120} $ & $ 0 $ & $ \frac{452348269}{15728640} $ & $ 0 $ & $ \frac{5851085490887}{251658240}$\\
\hline
\end{tabular*}}
\end{center}
\renewcommand{\arraystretch}{1.0}
\textbf{ Genus 3}
\renewcommand{\arraystretch}{1.15}
\begin{center}
{\scriptsize
\begin{tabular*}{0.9\textwidth}{@{\extracolsep{\fill}} | c | c c c c c | }
\hline
$n_2 \backslash n_1 $ & $ 2 $ & $ 4 $ & $ 6 $ & $ 8 $ & $ 10 $ \\ \hline
$0$ & $ 0 $ & $ \frac{6439}{6193152} $ & $ 0 $ & $ \frac{123167}{786432} $ & $ 0 $ \\
$1$ & $ -\frac{353}{1032192} $ & $ 0 $ & $ -\frac{724271}{24772608} $ & $ 0 $ & $ -\frac{342268673}{29360128} $ \\
$2$ & $ 0 $ & $ \frac{82823}{12386304} $ & $ 0 $ & $ \frac{468858317}{264241152} $ & $ 0 $ \\
$3$ & $ -\frac{2759}{1376256} $ & $ 0 $ & $ -\frac{41583137}{132120576} $ & $ 0 $ & $ -\frac{211129850593}{1056964608} $ \\
$4$ & $ 0 $ & $ \frac{416779}{6193152} $ & $ 0 $ & $ \frac{15342735559}{528482304} $ & $ 0 $ \\
$5$ & $ -\frac{914639}{49545216} $ & $ 0 $ & $ -\frac{3864359207}{792723456} $ & $ 0 $ & $ -\frac{2178379136683}{469762048} $ \\
$6$ & $ 0 $ & $ \frac{257963189}{264241152} $ & $ 0 $ & $ \frac{2719587683017}{4227858432} $ & $ 0 $ \\
$7$ & $ -\frac{48988931}{198180864} $ & $ 0 $ & $ -\frac{18042606251}{176160768} $ & $ 0 $ & $ -\frac{336935310613399}{2415919104}$\\
\hline
\end{tabular*}}
\end{center}
\renewcommand{\arraystretch}{1.0}
\textbf{ Genus 4}
\renewcommand{\arraystretch}{1.15}
\begin{center}
{\scriptsize
\begin{tabular*}{0.9\textwidth}{@{\extracolsep{\fill}} | c | c c c c c | }
\hline
$n_2 \backslash n_1 $ & $ 2 $ & $ 4 $ & $ 6 $ & $ 8 $ & $ 10 $ \\ \hline
$0$ & $ 0 $ & $ -\frac{2244757}{2477260800} $ & $ 0 $ & $ -\frac{283653643}{943718400} $ & $ 0 $ \\
$1$ & $ \frac{865427}{3715891200} $ & $ 0 $ & $ \frac{272614087}{5662310400} $ & $ 0 $ & $ \frac{11457822706721}{317089382400} $ \\
$2$ & $ 0 $ & $ -\frac{91054037}{9909043200} $ & $ 0 $ & $ -\frac{10649523253}{2202009600} $ & $ 0 $ \\
$3$ & $ \frac{9329603}{4246732800} $ & $ 0 $ & $ \frac{117628391911}{158544691200} $ & $ 0 $ & $ \frac{2091862017662453}{2536715059200} $ \\
$4$ & $ 0 $ & $ -\frac{590227019}{4404019200} $ & $ 0 $ & $ -\frac{4806828087037}{45298483200} $ & $ 0 $ \\
$5$ & $ \frac{1775895397}{59454259200} $ & $ 0 $ & $ \frac{4224848667521}{271790899200} $ & $ 0 $ & $ \frac{7773454487649391}{317089382400} $ \\
$6$ & $ 0 $ & $ -\frac{421624177657}{158544691200} $ & $ 0 $ & $ -\frac{7687488828890201}{2536715059200} $ & $ 0 $ \\
$7$ & $ \frac{116460407}{209715200} $ & $ 0 $ & $ \frac{76868168176019}{181193932800} $ & $ 0 $ & $ \frac{1772672261344760983}{1932735283200}$\\
\hline
\end{tabular*}}
\end{center}
\renewcommand{\arraystretch}{1.0}


\end{document}